\documentclass[preprint2]{aastex}

\usepackage{epsf}
\usepackage{graphicx}
\usepackage{color}

\def\eps@scaling{1.6}%

\newcommand\plotthree[3]{{%
 \typeout{Plotthree included the files #1 #2 #3}
 \centering
 \leavevmode
 \columnwidth=.30\columnwidth
 \includegraphics[width={\eps@scaling\columnwidth}]{#1}%
 \hfil
 \includegraphics[width={\eps@scaling\columnwidth}]{#2}%
 \hfil
 \includegraphics[width={\eps@scaling\columnwidth}]{#3}%
}}%

\newcommand\plotthreer[3]{{%
 \typeout{Plotthree included the files #1 #2 #3}
 \centering
 \leavevmode
 \columnwidth=.30\columnwidth
 \includegraphics[width={\eps@scaling\columnwidth}, angle=270]{#1}%
 \hfil
 \includegraphics[width={\eps@scaling\columnwidth}, angle=270]{#2}%
 \hfil
 \includegraphics[width={\eps@scaling\columnwidth}, angle=270]{#3}%
}}%

\newcommand\plotthreeone[1]{{%
 \typeout{Plotthree included the files #1 }
 \centering
 \leavevmode
 \columnwidth=.30\columnwidth
 \includegraphics[width={\eps@scaling\columnwidth}]{#1}%
}}%

\newcommand\plotthreetwo[2]{{%
 \typeout{Plotthree included the files #1 }
 \centering
 \leavevmode
 \columnwidth=.30\columnwidth
 \hfil
 \includegraphics[width={\eps@scaling\columnwidth}, angle=270]{#1}%
 \hfil
 \includegraphics[width={\eps@scaling\columnwidth}, angle=270]{#2}%
}}%

\newcommand{\be}{\begin{equation}}
\newcommand{\ee}{\end{equation}}

\newcommand{\ax}{$\alpha_{\rm X}$}
\newcommand{\auv}{$\alpha_{\rm UV}$}

\newcommand{\aox}{$\alpha_{\rm ox}$}

\newcommand{\rb}[1]{\raisebox{1.5ex}[-1.5ex]{#1}}
\newcommand{\msun}{$M_{\odot}$}

\newcommand{\plm}{$\pm$}

\newcommand{\swift}{{\it Swift}}

\newcommand{\chandra}{{\it Chandra}}

\newcommand{\lledd}{$L/L_{\rm Edd}$}
\newcommand{\erg}{ erg s$^{-1}$ cm$^{-2}$}
\shorttitle{Multi-wavelength observations of BAL QSOs}
\shortauthors{Grupe et al.}

\begin{document}

%\input DGrupe_clipfig.tex
%\useunitmm

\def\etal{{\it et\thinspace al.}\ }
\def\alp{{$\alpha$}\ }
\def\al2{{$\alpha^2$}\ }


\title{Is there a connection between Broad Absorption Line Quasars and Narrow
Line Seyfert 1 galaxies?
}


\author{Dirk Grupe\altaffilmark{1, 2},
%%S. Komossa\altaffilmark{3},
John.A. Nousek\altaffilmark{2}
%%Leisa Townsley\altaffilmark{2}
}

\altaffiltext{1}{Space Science Center, Morehead State University,
235 Martindale Dr., Morehead, KY 40351; d.grupe@moreheadstate.edu } 

\altaffiltext{2}{Department of Astronomy and Astrophysics, Pennsylvania State
University, 525 Davey Lab, University Park, PA 16802 }


%\altaffiltext{1}{Visiting Astronomer, Cerro Tololo Inter-American Observat}


\begin{abstract}
We consider whether Broad Absorption Line Quasars 
(BAL QSOs) and Narrow Line Seyfert
1 galaxies (NLS1s) are similar, 
as suggested by \citet{brandt00b} and \citet{boroson02}. 
For this purpose we constructed a sample
of 11 BAL QSOs from existing Chandra and Swift observations. 
We found that BAL QSOs and NLS1s both operate at high
Eddington ratios \lledd, although BAL QSOs have slightly lower \lledd. 
BAL QSOs and NLS1s in general have high
FeII/H$\beta$ and low [OIII]/H$\beta$ ratios following the classic 'Boroson \& Green' 
eigenvector 1 relation. 
We also
found that the mass accretion rates $\dot{M}$ of BAL QSOs and NLS1s are more similar than previously thought, 
although
some BAL QSOs exhibit extreme mass accretion rates of more than 10 \msun/year. 
These extreme mass accretion rates may suggest
that the black holes in BAL QSOs are relativistically
spinning.  Black hole masses in BAL QSOs are
a factor of 100 larger than NLS1s.  From their
location on a M-$\sigma$ plot, we find that BAL QSOs
contain fully developed black holes.  Applying
a principal component analysis to our sample
we find eigenvector 1 to correspond to the Eddington
ratio \lledd, and eigenvector 2  to black hole mass. 

\end{abstract}


\section{Introduction}

Outflows are a ubiquitous property of AGN. 
For example, blue-shifted emission lines 
like [OIII]$\lambda$5007 \citep[e.g., ][]{zhang11, komossa08},
or blue shifted absorption lines in the UV or X-rays \citep[e.g., ][]{crenshaw04} 
are typically
interpreted as signs of outflowing gas. 
Outflows can be driven in principle magnetically, thermally, and through radiation
\citep[e.g.,][]{kurasawa09a,kurasawa09b,proga04}.
Because of the large kinetic energy and angular
momentum transport outwards, outflows have 
strong influences on the AGN environment. As a consequence, many AGN parameters  
are driven by 
the Eddington ratio \lledd\ \citep[e.g., ][]{boroson02, sulentic00, grupe04, xu12}.
 The class of AGN that shows 
 the strongest outflows  as defined by 
 largest absorption column density and
 outflow velocity
 (besides jets) are Broad Absorption
Line Quasars \citep[BAL QSOs, e.g.,][]{weymann91} 
which can reach outflow velocities of 
 more than 20000 km s$^{-1}$ \citep[e.g.,][]{hamann08}.
Roughly 10-20\% of optically-selected 
quasars belong to this class \citep[e.g.,][]{dai08, elvis00} with 
even higher percentage
among infrared selected samples. 
However, it has also been suggested that the occurrence
of BALs may mark a specific time in the life of a quasar \citep{mathur00, becker00}.
Because the strength of radiation driven outflow directly
depends on \lledd, we infer that
BAL QSOs have extreme values for  \lledd
\citep[e.g.,][]{boroson02}. 

In the local Universe the AGN with
 the highest \lledd\ are Narrow Line Seyfert 1 galaxies \citep[NLS1s;][]{osterbrock85}, 
 which are historically defined as Seyfert 1s with a FWHM(H$\beta)<$2000 km s$^{-1}$ and [OIII]/H$\beta<3$ \citep{goodrich89}. 
Although this is a rather crude definition \citep[see for example][]{zamfir09, marziani09},
 for the purpose of our study we adopt this definition throughout the paper. 
NLS1s have drawn a lot of attention over the last two decades due to their extreme 
properties, such as, on average, steep X-ray spectra, strong Fe II emission and weak
emission from the Narrow Line Region \citep[e.g.,][]{boroson92, grupe04, komossa08r}. 
All these properties are linked, 
and are most likely driven primarily by the mass of the central black hole and 
the Eddington ratio \lledd.  Generally speaking, NLS1s are  AGN with 
low black hole masses and 
high \lledd.
NLS1s have also been considered to be AGN in an
early stage of their evolution \citep[][]{grupe99, mathur00}.

It has been suggested by \citet{brandt00b} and \citet{boroson02}
that BAL QSOs and NLS1s are similar with respect to their 
high \lledd. It has also been found that the rest-frame
optical spectra of at least some
BAL QSOs look very much like low-redshift NLS1s 
\citep[e.g.,][]{marziani09, dietrich09}.
As pointed out by \citet{boroson02}, BAL QSOs and NLS1s 
have very similar FeII/H$\beta$ and [OIII]/H$\beta$ ratios and
follow the classical '\citet{boroson92} eigenvector 1' relation.
This is remarkable, because BAL QSOs and NLS1s
differ in their black hole masses and 
appear to be at opposite ends of the $M_{\rm BH}$
spectrum. Typically Seyfert 1s with large black hole masses have larger 
[OIII]/H$\beta$ and smaller FeII/H$\beta$ ratios than NLS1s (or BAL QSOs) as shown 
for example in \citet{grupe04} and \citet{grupe10}.  Although there seem to
be many similarities between NLS1s and BLA QSOs, as pointed out by
\citet{laor02}, the strengths of the broad absorption lines (BALs) 
 seem to be directly correlated with the luminosity of the AGN and BALs
appear only in highly-luminous AGN.

So far only one object has been established that shows a clear
connection between
NLS1s and BAL QSO:
the X-ray transient NLS1 WPVS 007 \citep[e.g.][]{grupe13}. This NLS1s was discovered as a
bright X-ray AGN during the ROSAT All-Sky Survey \citep[RASS,][]{voges99}, but showed a
dramatic drop in its X-ray flux when observed a few years later \citep[][]{grupe95}. Various
follow-up observations by ROSAT, Chandra, XMM and \swift\ all confirmed this X-ray low state
\citep[][and references therein]{grupe08b, grupe13}. The X-ray spectra of WPVS 007 can be
modeled by a power law with a strong partial covering absorber in the line of sight
\citep[][]{grupe08b, grupe13}.  UV spectroscopy by FUSE and HST revealed that strong board
absorption line features had evolved within just a decade  \citep[][,and Cooper et al. 2014, in prep]{leighly09, cooper13}. 

Another NLS1 suggested to be a link between NLS1s and BAL QSOs is Mkn 335 which was discovered
by \swift\ in May 2007 to be in a deep X-ray flux state \citep{grupe07b}. Although the X-ray
data of Mkn 335 can be modeled by a partial covering absorber \citep{grupe08, grupe12} they can
also be fitted by reflection models \citep{grupe08, gallo13}. 
Nevertheless, Mkn 335 has also
developed UV absorption lines as reported by \citep{longinotti13}. This new
finding may suggest that Mkn 335 will develop similar UV absorption lines 
as WPVS 007 did over the last two decades.

Although BAL QSOs were
considered  not to be variable, more recent studies have shown that BAL QSOs
indeed show variability not only in their UV absorption lines 
\citep[e.g.]{filiz12, filiz13, hamann08, capellupo11, capellupo12}, 
 also in X-rays as reported by \citet{saez12}.
On the other hand, NLS1s are known to be highly variable in 
X-rays \citep[e.g.][]{grupe01, grupe10}.

The goal of this paper is to search for similarities in the spectral energy distributions and
emission line properties of BAL QSOs and NLS1s. The motivation is to test if NLS1s and
BAL QSOs are both high \lledd\ AGN but appear to have different distributions of their black
hole masses and mass accretion rates, as suggested by \citet{brandt00b} and \citet{boroson02}. 
The outline of this paper is as follows: in \S\,\ref{observe} we describe 
sample selection and the 
observations and data reduction by \swift\ and \chandra, as well as 
in the infrared and in the optical.  In \S\,\ref{results} we 
present the results from the analysis of the spectral energy distributions.
In \S\,\ref{discuss} we discuss the results. 
Throughout the paper spectral indices are denoted as energy spectral indices
with
$F_{\nu} \propto \nu^{-\alpha}$. Luminosities are calculated assuming a $\Lambda$CDM
cosmology with $\Omega_{\rm M}$=0.27, $\Omega_{\Lambda}$=0.73 and a Hubble
constant of $H_0$=75 km s$^{-1}$ Mpc$^{-1}$. 
Luminosity distances were estimated 
using the cosmology calculator  by \citet{wright06}. 
All errors are 1$\sigma$ unless stated otherwise. 
{Note that although in recent years slightly lower values of $H_0$ have been reported, in
particular from the {\it Planck} measurements \citep{ade14}, we continue to use $H_0$ =
75 km s$^{-1}$ Mpc$^{-1}$ in order to allow a direct comparison with luminosities of our
other samples. The differences in luminosity between the two values are on the order of
20\%.}
 
\section{\label{observe} Observations and data reduction}

\subsection{Sample Selection \label{sample}}
 BAL QSOs generally appear to be X-ray
weak following the definitions by \citet{brandt00} and \citet{gibson09}. 
Therefore, compared with other classes
 of AGN, X-ray observation of BAL QSOs appear to be rather sparse and only a small number of
 known BAL QSOs have been followed up in X-rays \citep[e.g.][]{grupe03, saez12}.
 Usually this X-ray weakness is explained by 
strong intrinsic absorption in X-rays, typically of the order of several
$10^{22}$ cm$^{-2}$ \citep[e.g.][]{grupe03, saez12}. However, some BAL QSO, like PG
1004+130 and PG 1700+518 may be intrinsically X-ray weak as pointed out by \citet{luo13}. 
As recently reported by \citet{luo14}, observations by NuStar revealed that intrinsic
X-ray weakness seem to be quite common among BAL QSOs. The \citet{luo14} sample also
contains several of BAL QSOs discussed in our paper: besides PG 1004+130 and PG 1700+518,
as mentioned above, also IRAS 07598+6508, PG 0946+301, PG 1001+054, Mkn 231, and IRAS
14026+4341.
In other
words, these are BAL QSO analogies of PHL 1811 \citep[e.g.][]{leighly07}.
 The best-suited X-ray
observatory to at least detect a BAL QSO in X-rays is \chandra\ due its superior imaging
qualities. On the other hand, the most efficient Optical/UV instrument in space is the
UV-Optical telescope \citep[UVOT,][]{roming05} onboard \swift\
\citep{gehrels04}. The aim of
this study is to combine these two missions to obtain spectral energy 
distributions for BAL
QSOs which then can be compared with already existing X-ray and UV/Optical
data from \swift\ observations of NLS1s.
The \swift\ observations of most of the NLS1s have been  already published  in
 \citet{grupe10}. We cross-correlated the catalogue of \chandra\ observations of BAL QSOs 
performed as part of the Penn
State ACIS-S Guaranteed Time Program with the \swift\ master observation catalogue. An additional
requirement was that optical spectra were publicly available. Out of 306 sources that have
\chandra\ PSU GTO time (PI G. Garmire)
and \swift\ observations we found 11 BAL QSOs for which we also found
existing optical spectroscopy data as listed in Table\,\ref{chandra_log}.
The \chandra\ observations of the majority of these
BAL QSOs were already published by \citet{saez12}. 
The three sources not listed
in \citet{saez12} are SDSS J073733+392037, PG1115+080, and SDSS J143748+432707.

\subsection{\chandra\ Observations}

The coordinates, redshift, luminosity distance, Galactic column density, reddening, and 
Chandra observing parameters of the 11 BAL QSOs are 
summarized in Table\,\ref{chandra_log}. The primary and secondary data were retrieved from
the \chandra\ archive. 
All data analysis was done with CIAO version 
4.6 using the most recent calibration data base 4.5.9.
Source counts were extracted within a circle with a radius of 1.5$^{"}$ and background counts
in a nearby source-free circular region with a radius or 15$^{"}$. We applied the CIAO tasks
{\it mkrmf} and {\it mkarf} to create the appropriate redistribution matrix and ancillary
response files, respectively. The X-ray spectra were analyzed using XSPEC version 12.7.1
\citep{arnaud96}. For the majority of spectra we applied Cash statistics \citep{cash79}
 to perform the fits in XSPEC. Because the focus of this paper is on constructing the
spectral energy distributions we only performed a simple spectral analysis of the X-ray data. 
A more detailed analysis of the \chandra\ spectra can be found in \citet{saez12}.
Here, we perform a homogeneous re-analysis of all the sources presented
in this paper, including fits with intrinsic absorbers, such as a
 partial covering absorber.

\subsection{\swift\ Observations}

Table\,\ref{swift_log} presents the 
\swift\ observations  including the start and end times and the total exposure
times. 
The \swift\ X-ray telescope \citep[XRT;][]{burrows05} 
was operating in photon counting mode \citep{hill04} and the
data were reduced by the task {\it xrtpipeline} version 0.12.6., 
which is included in the HEASOFT package 6.12. Source counts were selected in a
circle with a radius of 24.8$^{''}$ and background counts in a nearby 
circular region with a radius of 247.5$^{''}$. 
The X-ray spectra were analyzed using {\it XSPEC} version 12.7.1
     \citep{arnaud96}.

The UV-Optical Telescope \citep[UVOT;][]{roming05} 
data of each segment were coadded in each filter with the UVOT
task {\it uvotimsum}.
Source counts in all 6 UVOT filters
  were selected in a circle with a radius of 5$^{''}$ and background counts in
  a nearby source free region with a radius of 20$^{''}$.
  UVOT magnitudes and fluxes were measured with the task {\it  
uvotsource} based on the most recent UVOT calibration as described in  
\citet{poole08} and  \citet{breeveld10}.
The UVOT data were corrected for Galactic reddening
\citep[][]{sfd98}. The correction factor in each  
filter was
calculated with equation (2) in \citet{roming09}
who used the standard reddening correction curves by \citet{cardelli89}.

Although \swift\ observations existed for
most of these BAL QSOs, for 5 of these sources we requested new observations to obtain UVOT
observations in all 6 UVOT filters. All these additional observations were performed in
January and June 2014 and were typically of the order of 2ks as listed in 
Table\,\ref{swift_log}.

\subsection{Infrared data}
In order to extend the spectral energy distributions to lower energies we obtained publicly
available data in the near infrared from the Two Micron All-Sly Survey
\citep[2MASS,][]{skrutskie06} and in the
mid-infrared from the {\it Wide-field Infrared Survey Explorer} mission {\it WISE}
\citep{wright10}. These infrared data were retrieved from the public archive at the  
Infrared Processing and Analysis Center using the online GATOR tool. 
The observed fluxes in the {\it 2MASS} J, H, K, and {\it WISE} W1, W2, W3, and W4 bands are
listed in Table\,\ref{ir_res}. 

\subsection{Optical Spectroscopic data \label{opt_dat}}
Optical spectra of the BAL QSOs were primarily derived from the Sloan Digital Sky Survey
\citep[SDSS][]{york00} Data Release 10 \citep[DR10,][]{ahn13}, except for IRAS 07598+6508, Mkn
231, PG 1700+518, and PG 2112+059. 
For these objects we used the optical spectra published in \citet{hines95},
\citet{moustakas06}, and \citet{boroson92}, respectively.
 For all measurements of optical line properties
we subtracted an FeII template which was based on the \citet{boroson92} template, 
as described in \citet{grupe04a}. For the black hole estimates we applied the relations given in 
\citet{vestergaard06} and \citet{vestergaard09}.

\section{\label{results} Results}

\subsection{Analysis of the \chandra\ data}
All Chandra data were fitted initially with an absorbed power law model with the Galactic
absorption column density $N_{\rm H}$ listed in Table\,\,\ref{chandra_log}. While
this model seems appropriate for most objects, 4 of the sources require an
intrinsic absorber at the redshift of the source. While PG 0946+301 can be fitted
by a simple neutral absorber, the other three BAL QSOs, PG 1004+130, PG1115+080,
and Mkn 231 need be fitted by a partial covering absorber model. The results of all
fits to the Chandra data are summarized in Table\,\ref{chandra_res}, including  
the X-ray fluxes in the 0.3-10 keV observed frame. In addition, Table\ref{chandra_res} also lists the
 Optical to X-ray spectral slopes \aox.
  These were determined based 
on the spectral energy distributions using
the \chandra\ and \swift\ data as shown in  Figure\,\ref{sed_plots} 
(see Section\,\ref{seds}).
 Note, however, that the spectral models used here are just a phenomenological description of the
spectrum and do not necessarily represent the underlying physics in the source. For example, the
geometry of Mkn 231 is highly complex consisting of absorption components as well as contributions
from X-ray emission from strong surrounding starburst regions \citep[e.g.][,and references
therein]{teng14}. Nevertheless, our results obtained for PG 1004+130 agree well with those found
from the combined \chandra\ and NuStar data in \citet{luo13}.

\subsection{Spectral Energy Distribution \label{seds}}
\begin{figure*}
%\epsscale{0.75}
\epsscale{1.5}
\plotthreer{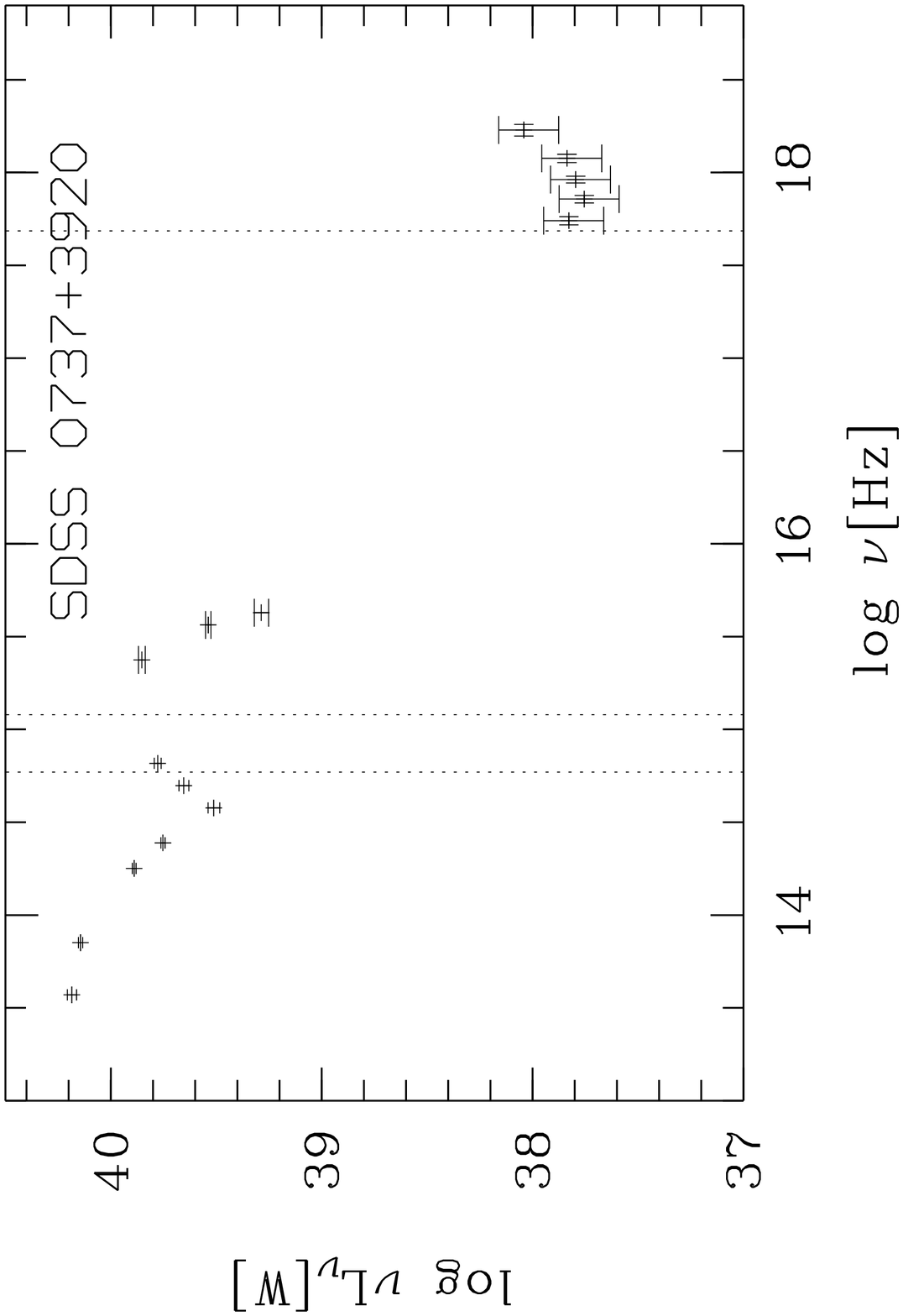}{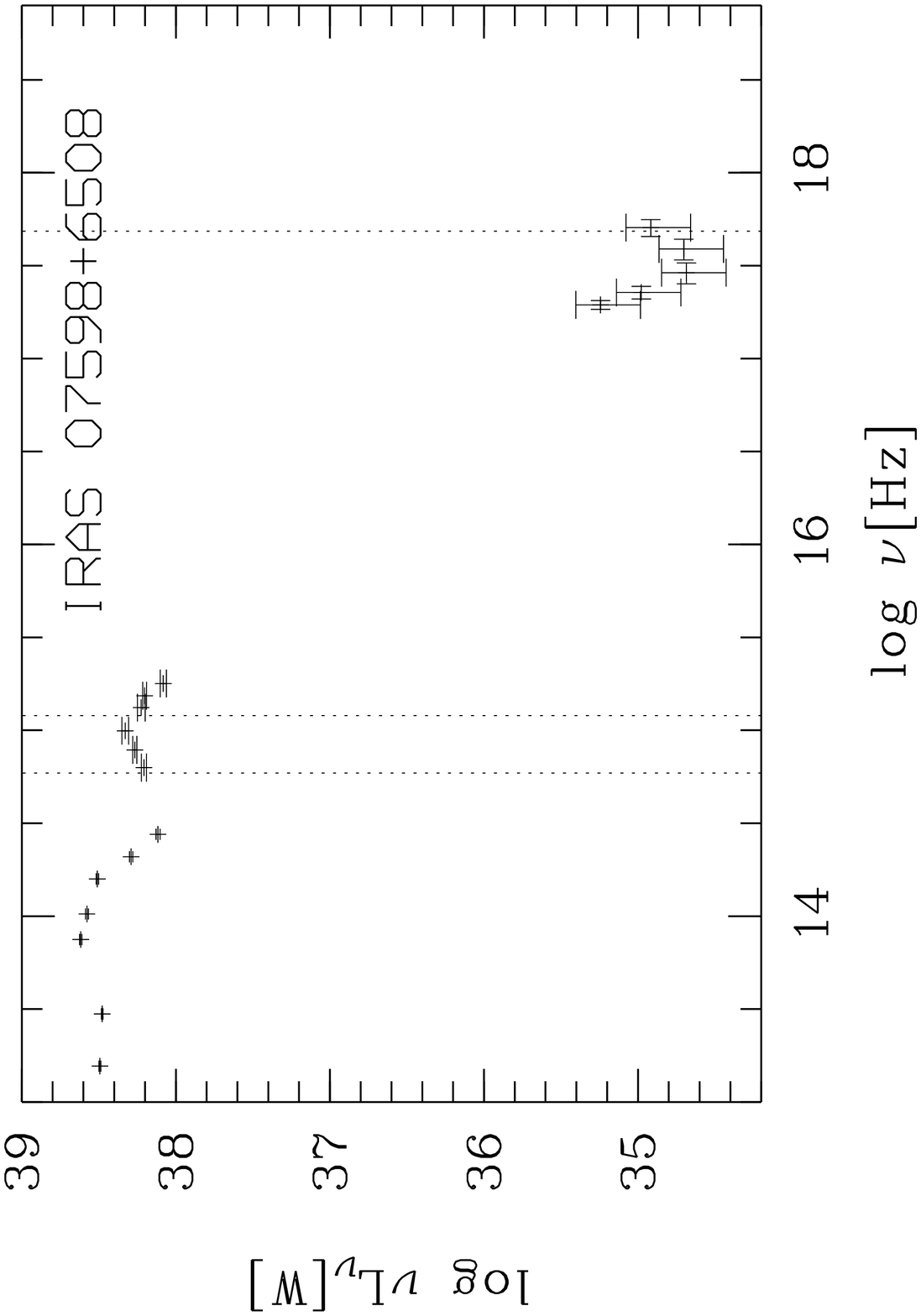}{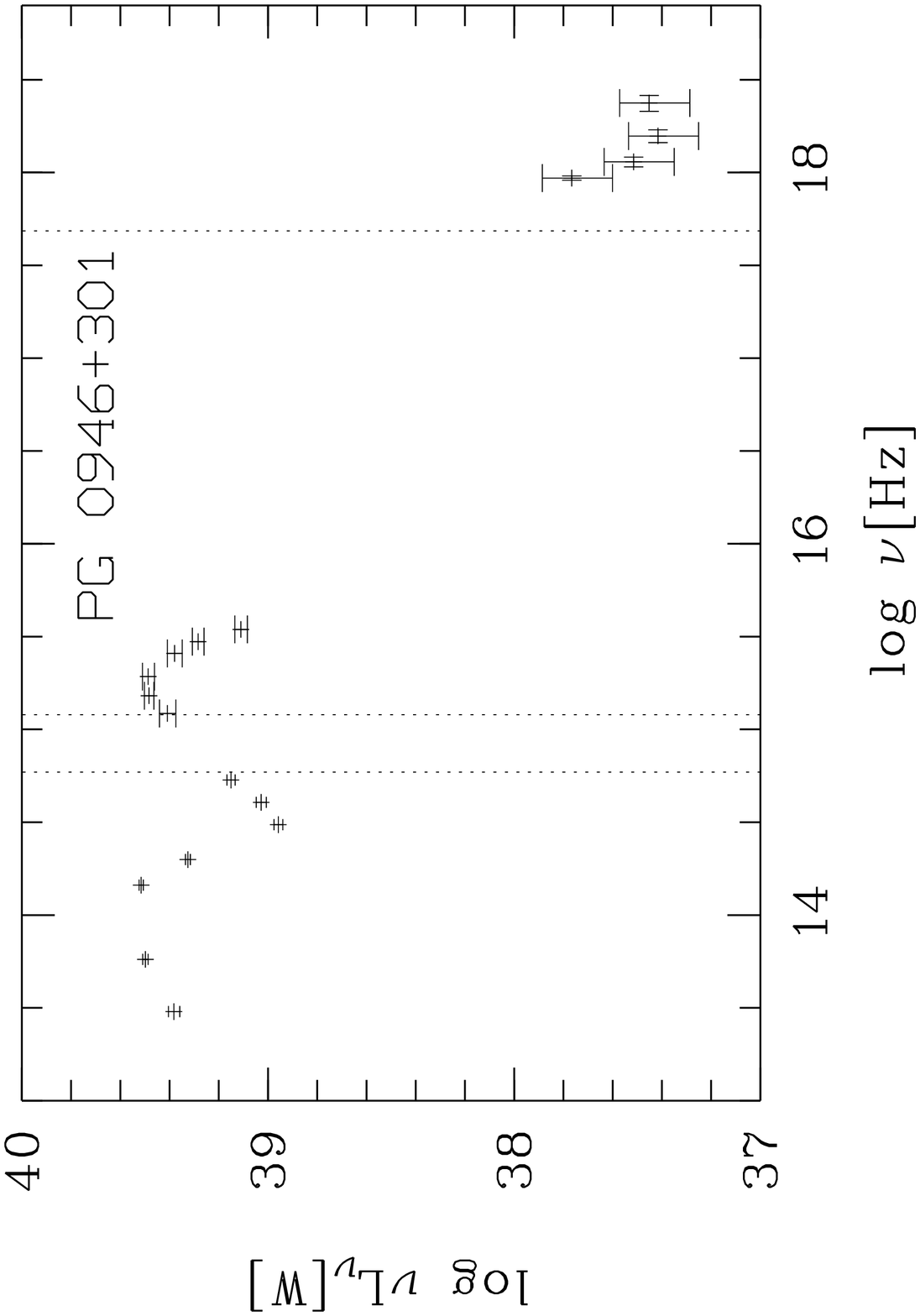}

\plotthreer{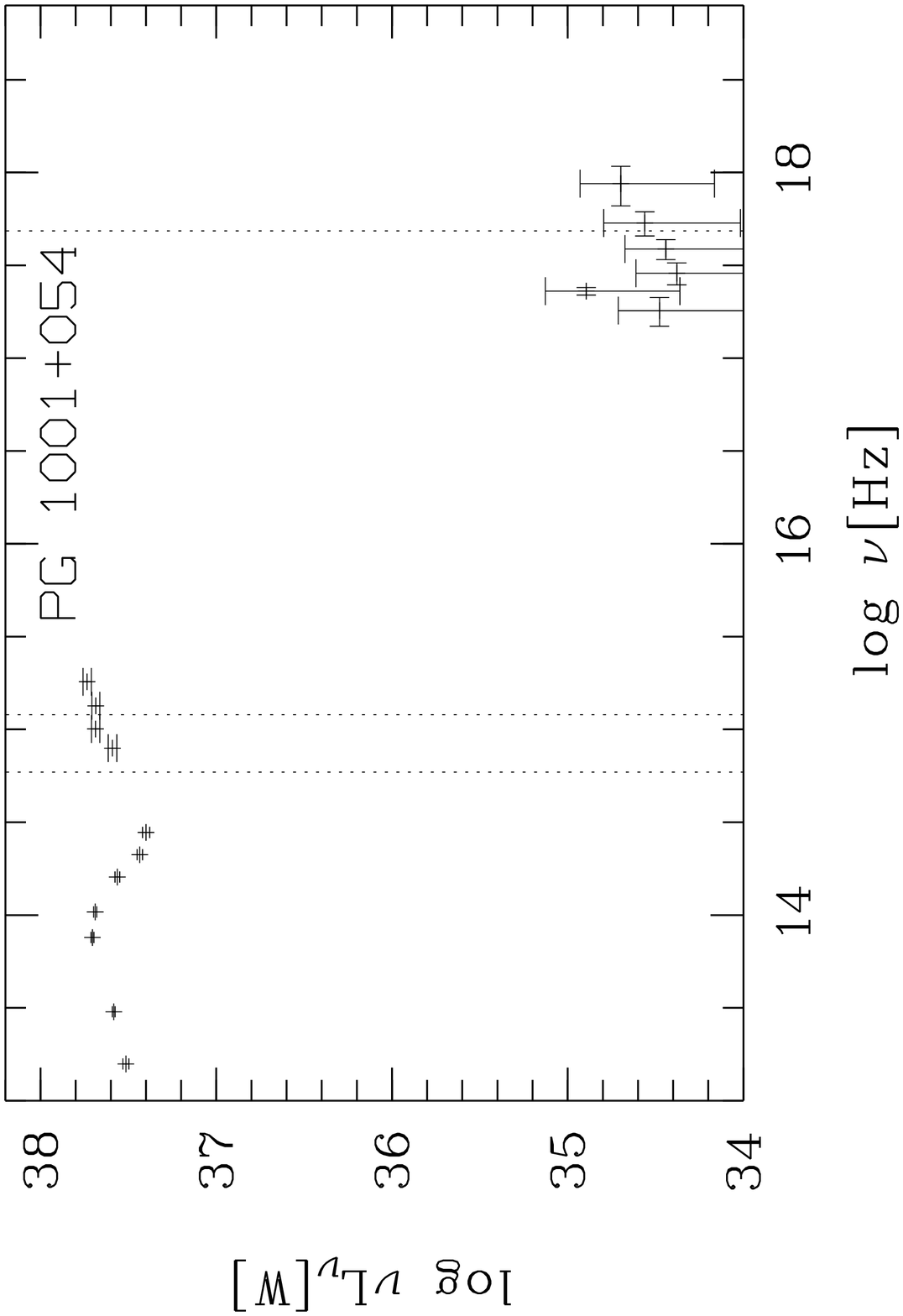}{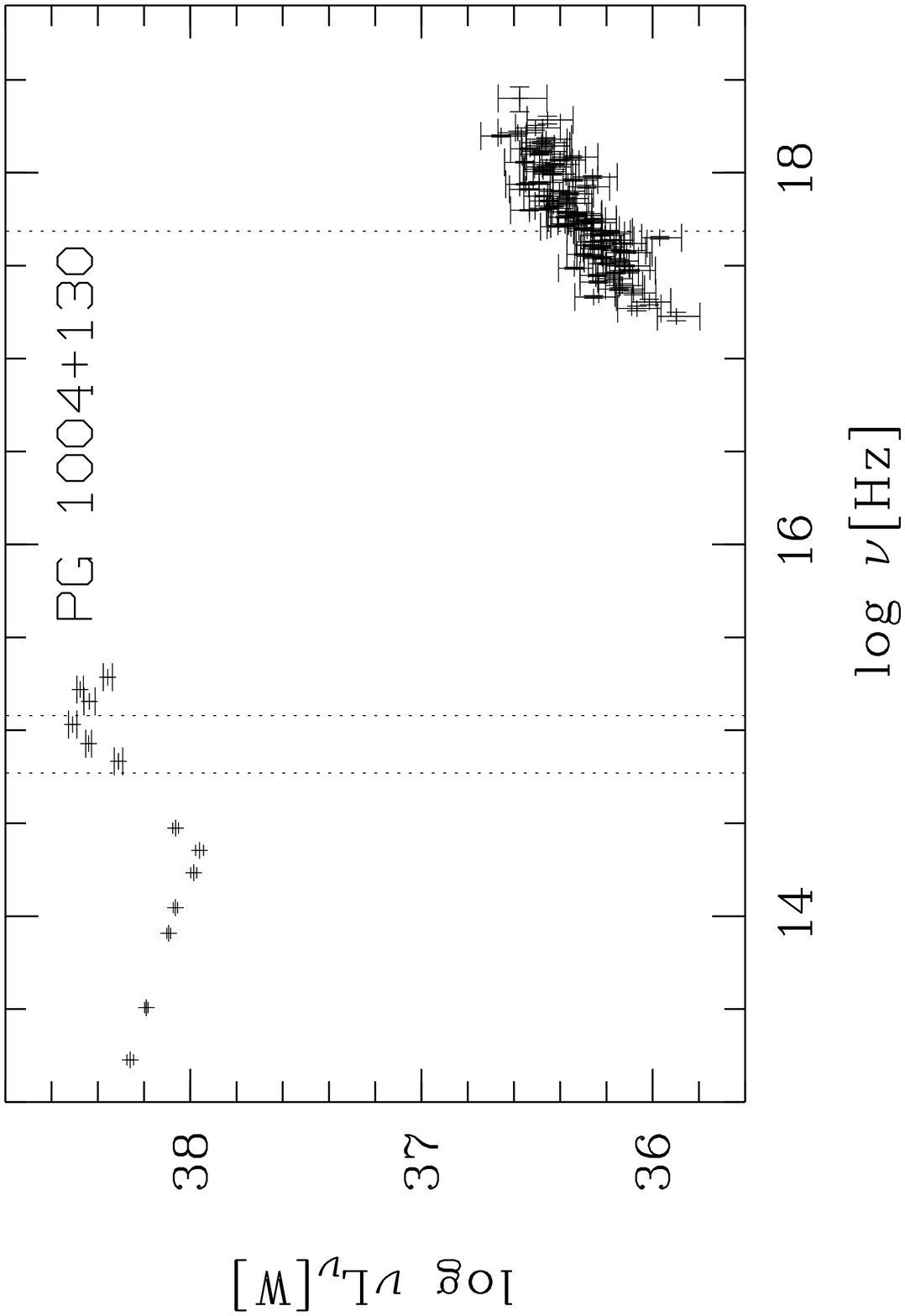}{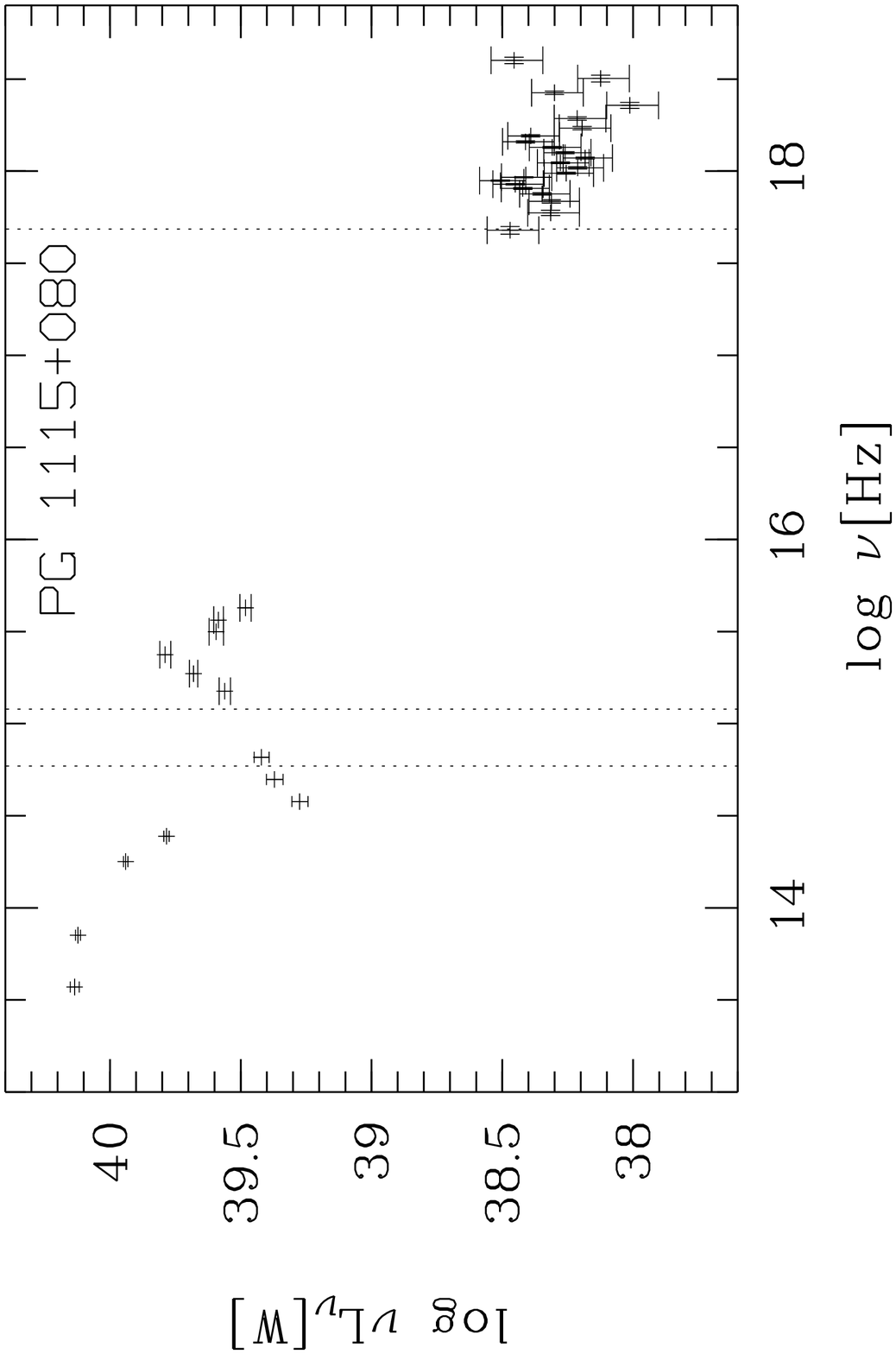}

\plotthreer{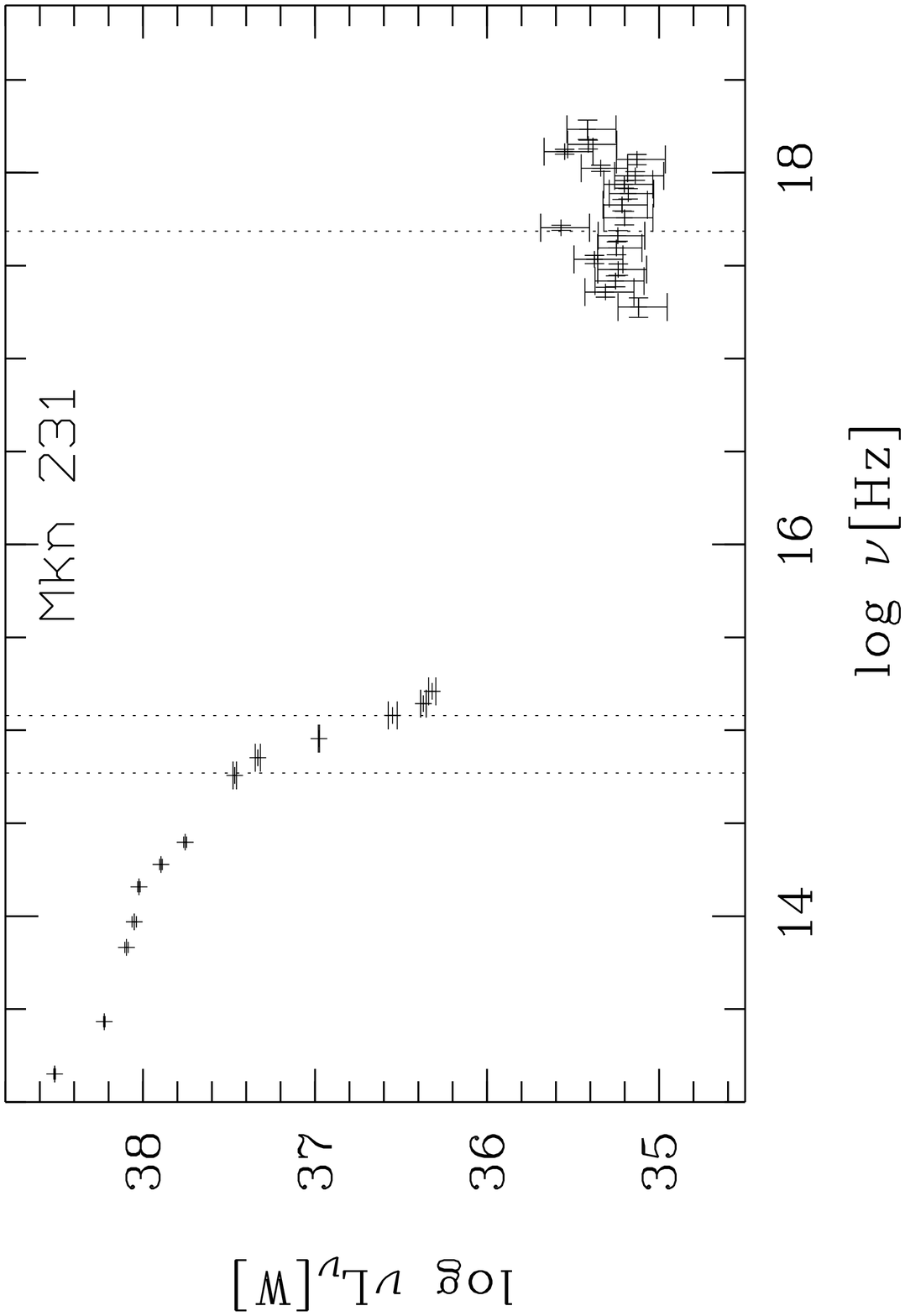}{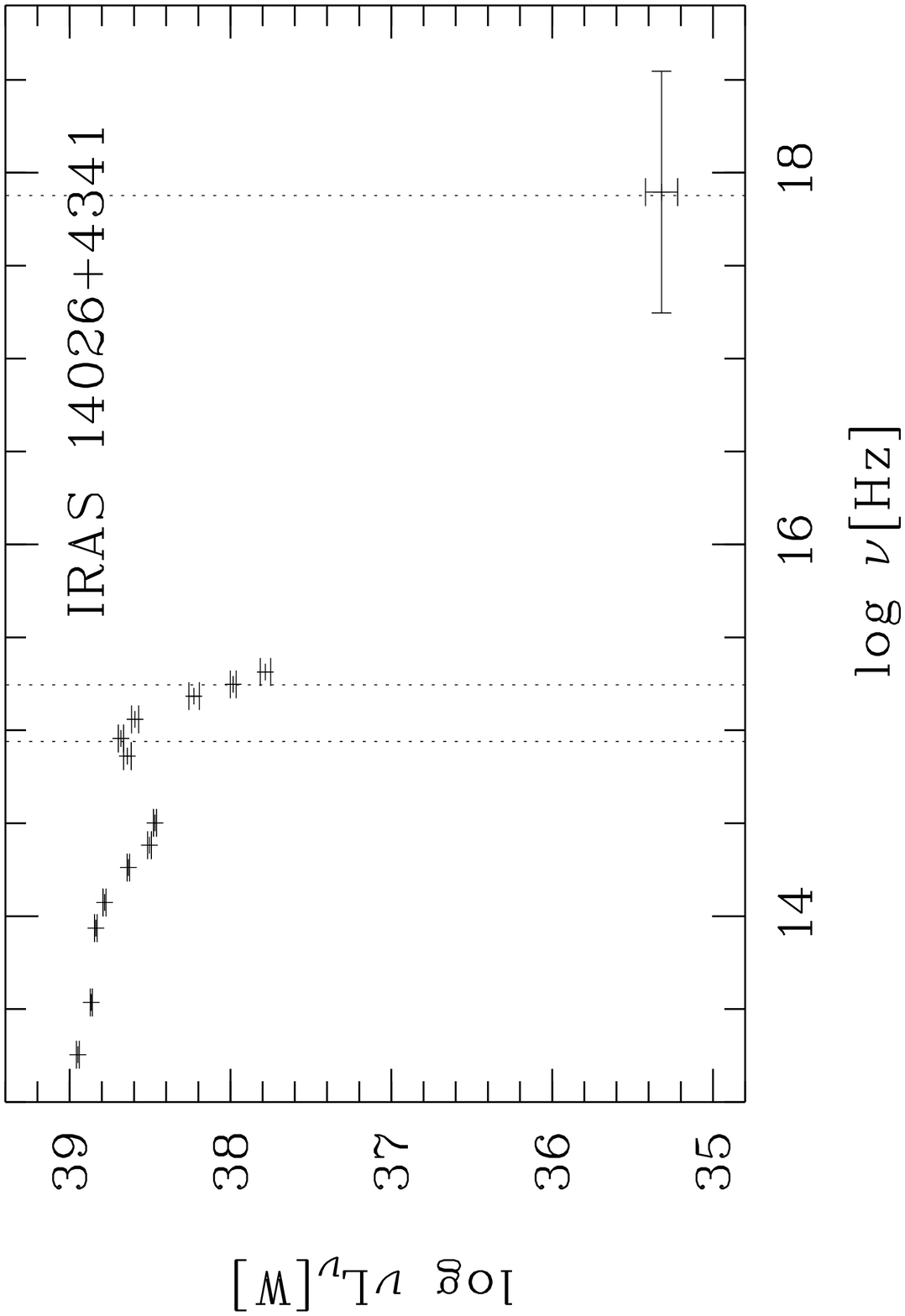}{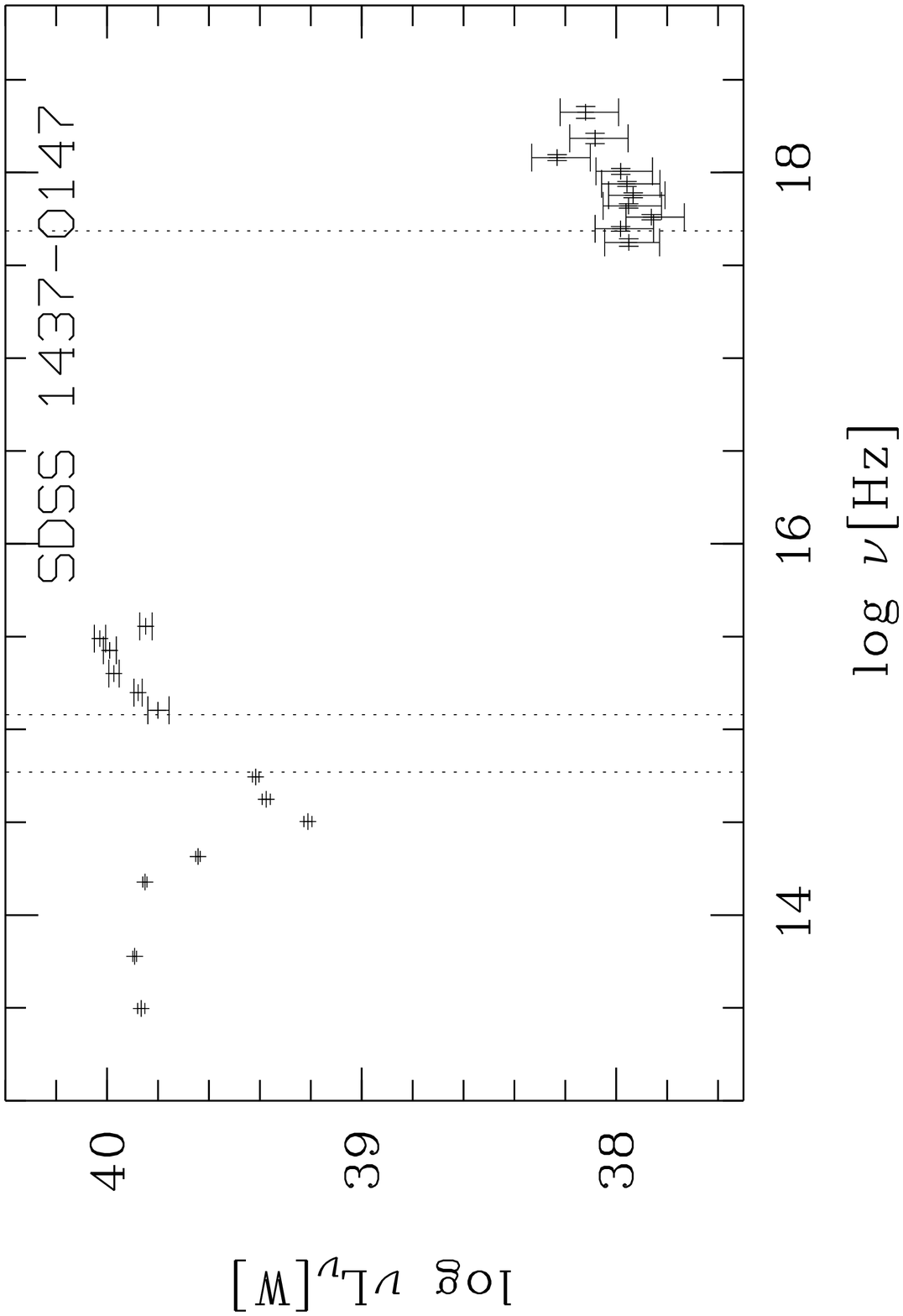}

\plotthreetwo{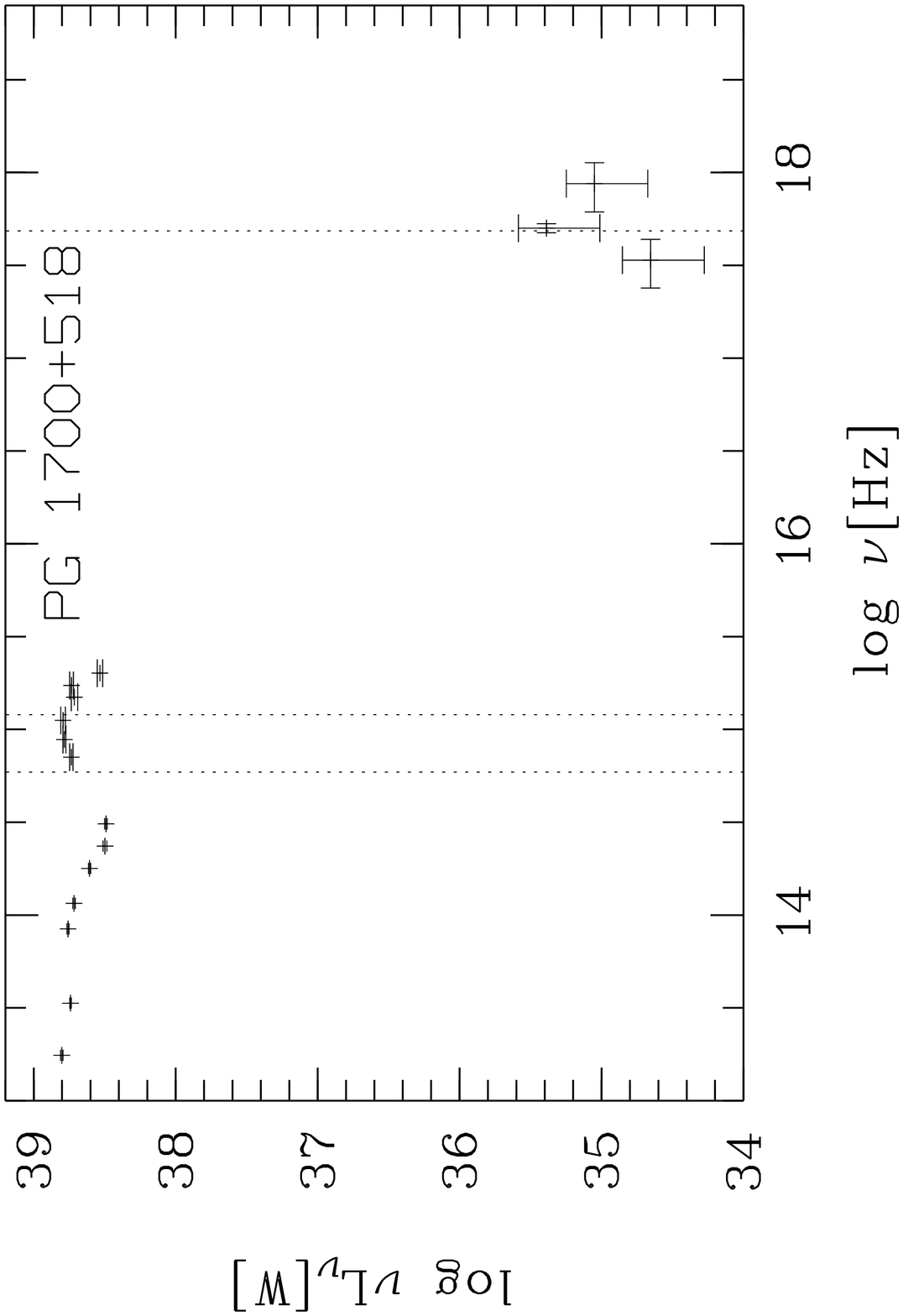}{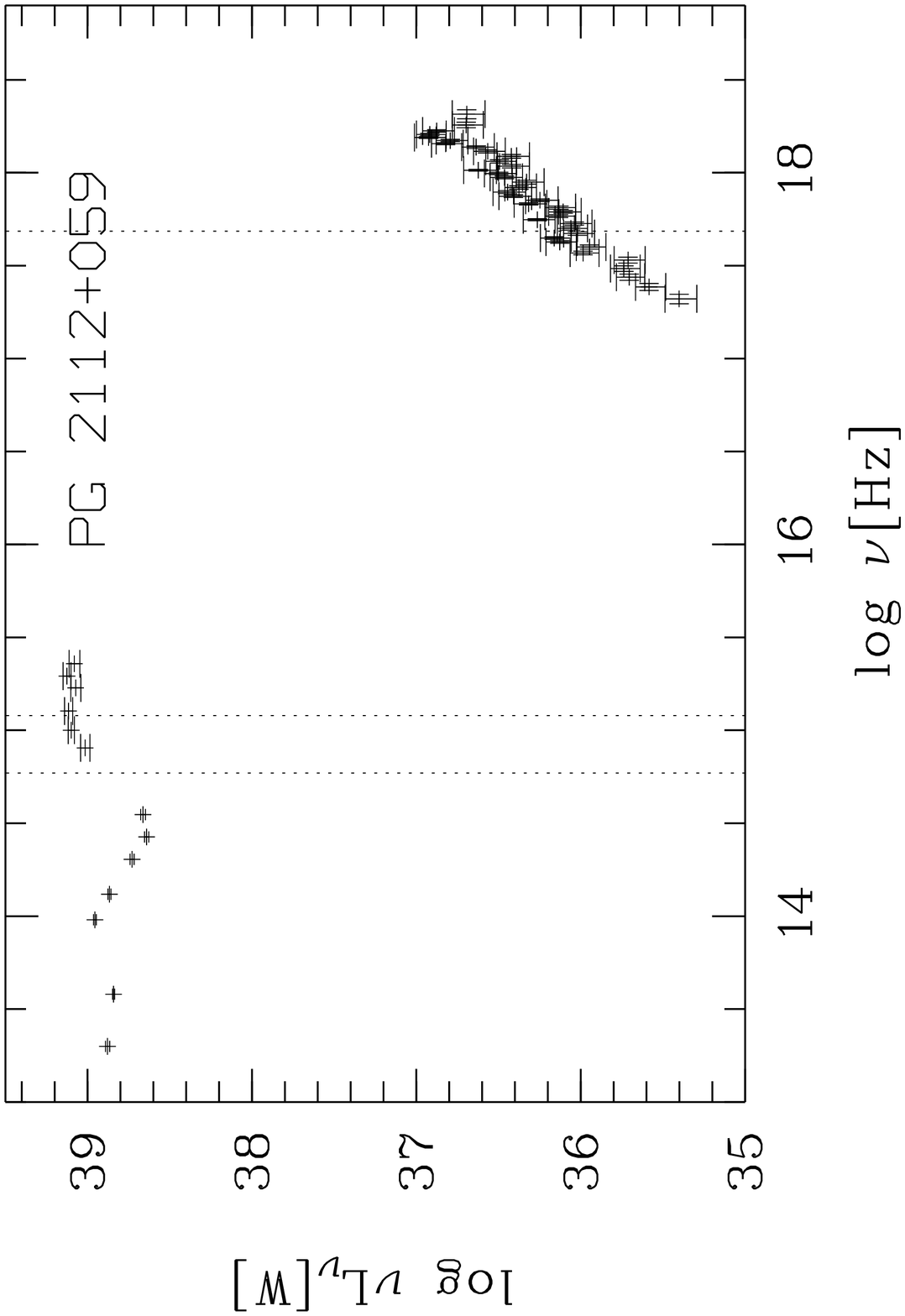}
\caption{\label{sed_plots} Rest-frame k-corrected IR to X-ray Spectral Energy Distributions of the BAL QSOs.
The vertical dashed lines make the rest-frame 5500\AA, 2500\AA, and 2keV points. \swift\  UVOT and
XRT data have been corrected for Galactic reddening and absorption. 
}
\end{figure*}

The spectral energy distributions (SEDs) displayed in Figure\,\ref{sed_plots}
were constructed by using the WISE, 2MASS, \swift\ UVOT, and \chandra\ X-ray data.
All SEDs are shown in rest-frame with k-corrected luminosities\footnote{Applying the k-correction as
 defined by \citet{oke68}
with $L_{\rm rest} = L_{\rm obs} \times (1+z)^{\alpha - 1}$}.
 Note that for IRAS
14026+4341 the number of counts detected during the 1.5 ks \chandra\ observation was 
too low to obtain a spectral fit. We therefore only give one data point in the 0.3-10 keV
regime in the SED which was determined from the count rate during the \chandra\ observation
and converted to flux units assuming a
standard AGN spectrum with \ax=1.0 and Galactic absorption.
The bolometric luminosities $L_{\rm bol}$ were measured by integrating over the k-corrected SEDs
as shown in 
Figure\,\ref{sed_plots} between 1$\mu$m and 10 keV (10$^{14}$ - $2\times 10^{18}$ Hz). These bolometric 
luminosities are listed in Table\,\ref{mbh_lledd}. The SEDs were also used to determine \aox. 
These Optical to X-ray spectral slopes are listed in Table\,\ref{chandra_res} together with the expected value
estimated from the k-corrected rest-frame luminosity density at 2500\AA\ applying equation (12)  in
\citet{grupe10}. Four of the 11 BAL QSOs appear to be clearly X-ray weak following the definition by 
\citet{brandt00} with \aox$>$2.0. 
 The \aox\ criterion for X-ray weakness, however, depends on luminosity.
 We also determined the difference
between the measured \aox\ and the expected value for the Optical-to-X-ray 
spectral slope $\alpha_{\rm ox, expected}$ following the relation
given in \citet{grupe10}. As described in \citet{gibson09} this parameter 
is defined as
$\Delta$\aox\ = \aox - $\alpha_{\rm ox, expected}$
and for the BAL QSOs in our sample is listed in Table\,\ref{chandra_res}.  

Although for BAL QSOs
absorption is the first thought when a source is detected to be X-ray weak \citep[e.g.][]{grupe03}, 
spectral analysis of the data for these four source does not suggest the presence of a strong absorber.
Interestingly, the BAL QSOs in which an intrinsic absorber is clearly detected appear not to be X-ray weak.
Alternatively, these sources could be heavily absorbed (i.e. Compton thick), and
  only a  reflected fraction of the intrinsic emission is actually seen.
  Also, given the quality of the spectral data, a partial covering geometry
  cannot be excluded. In a partial covering absorber scenario 
  only  a fraction of the intrinsic continuum
  emission is seen directly, while the rest is strongly obscured.
Also note again, that some of the BAL QSOs, like PG 1004+130 and PG 1700+518 may be intrinsically X-ray weak
\citep{luo13}.

\subsection{Statistical Analysis}
If BAL QSOs and NLS1s  represent related phenomena, then some of their characteristic intrinsic
 properties should be similar.
 For our study we use the standard definition of
NLS1s with a cut off line at FWHM(H$\beta$)=2000 km s$^{-1}$. 

\subsubsection{Distributions}
\begin{figure*}
%\epsscale{0.75}
\epsscale{1.5}
\plotthreer{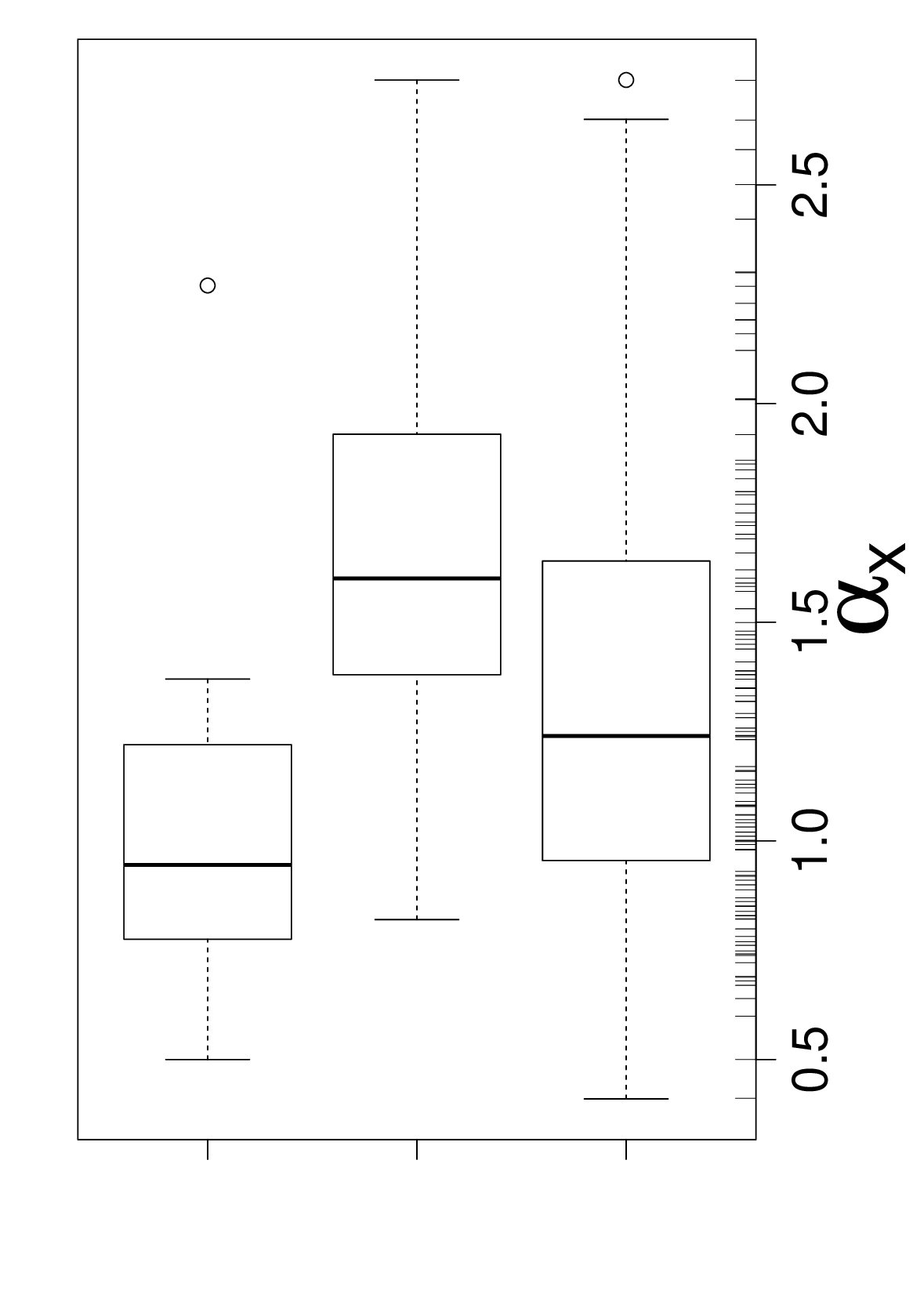}{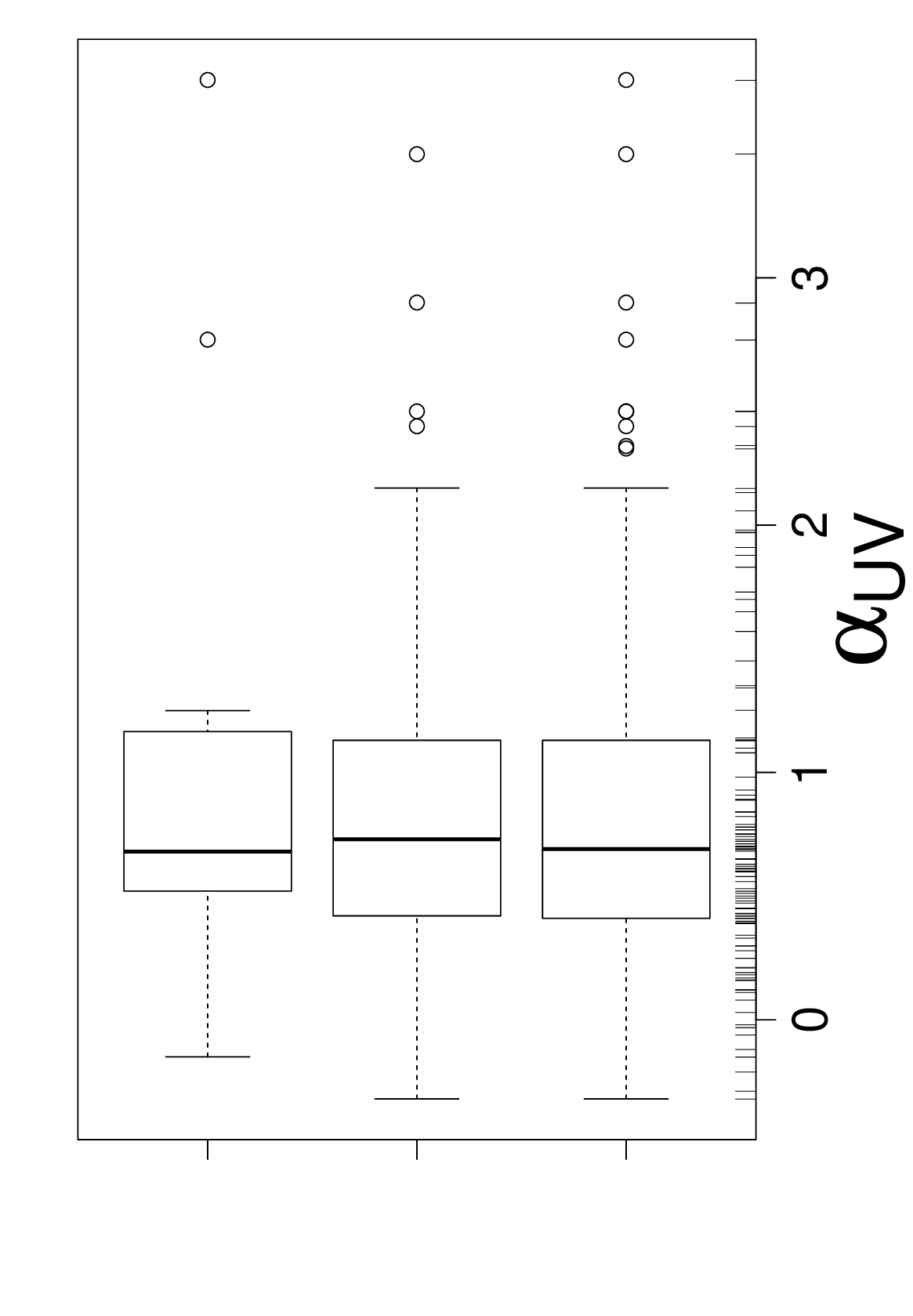}{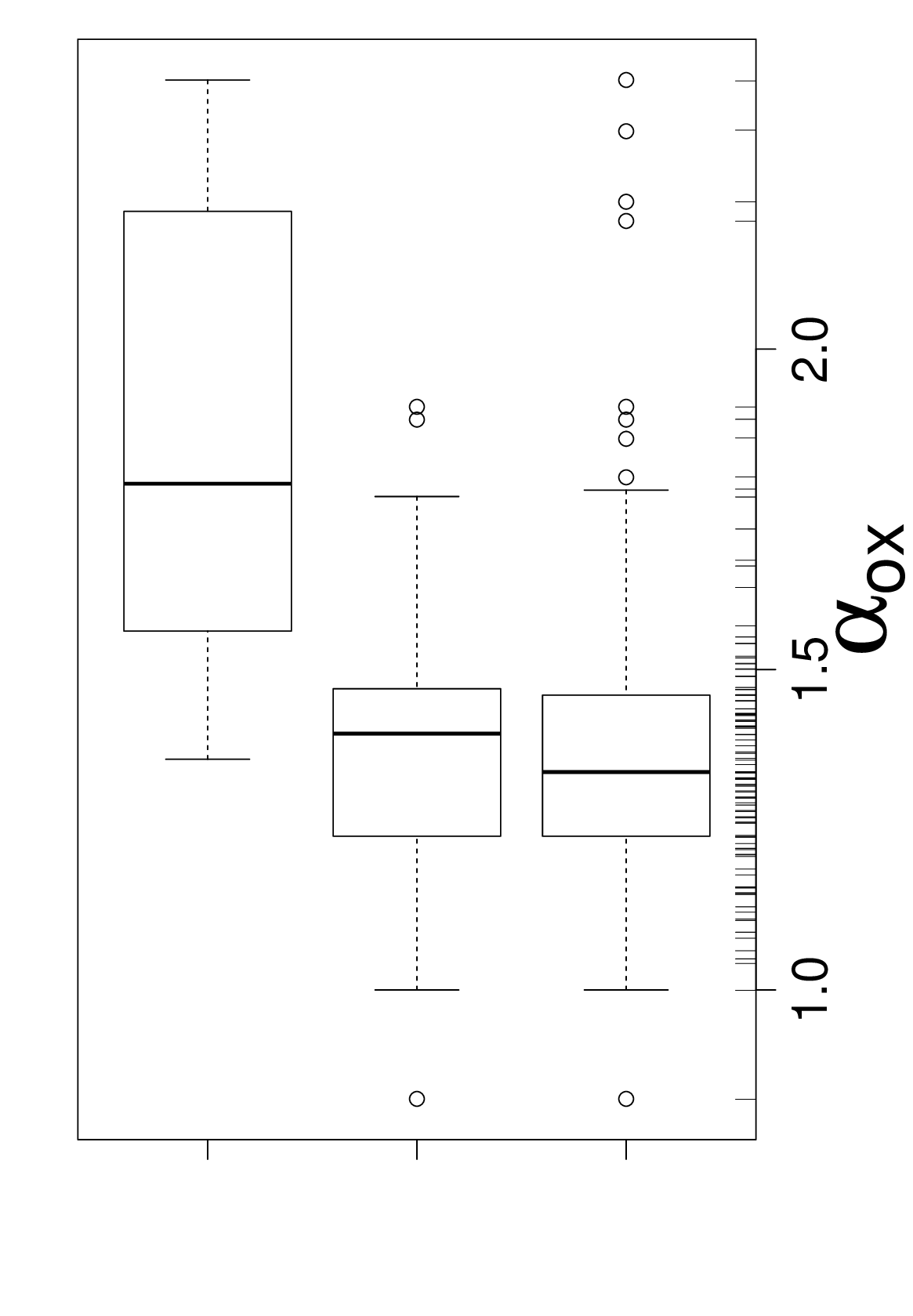}

\plotthreer{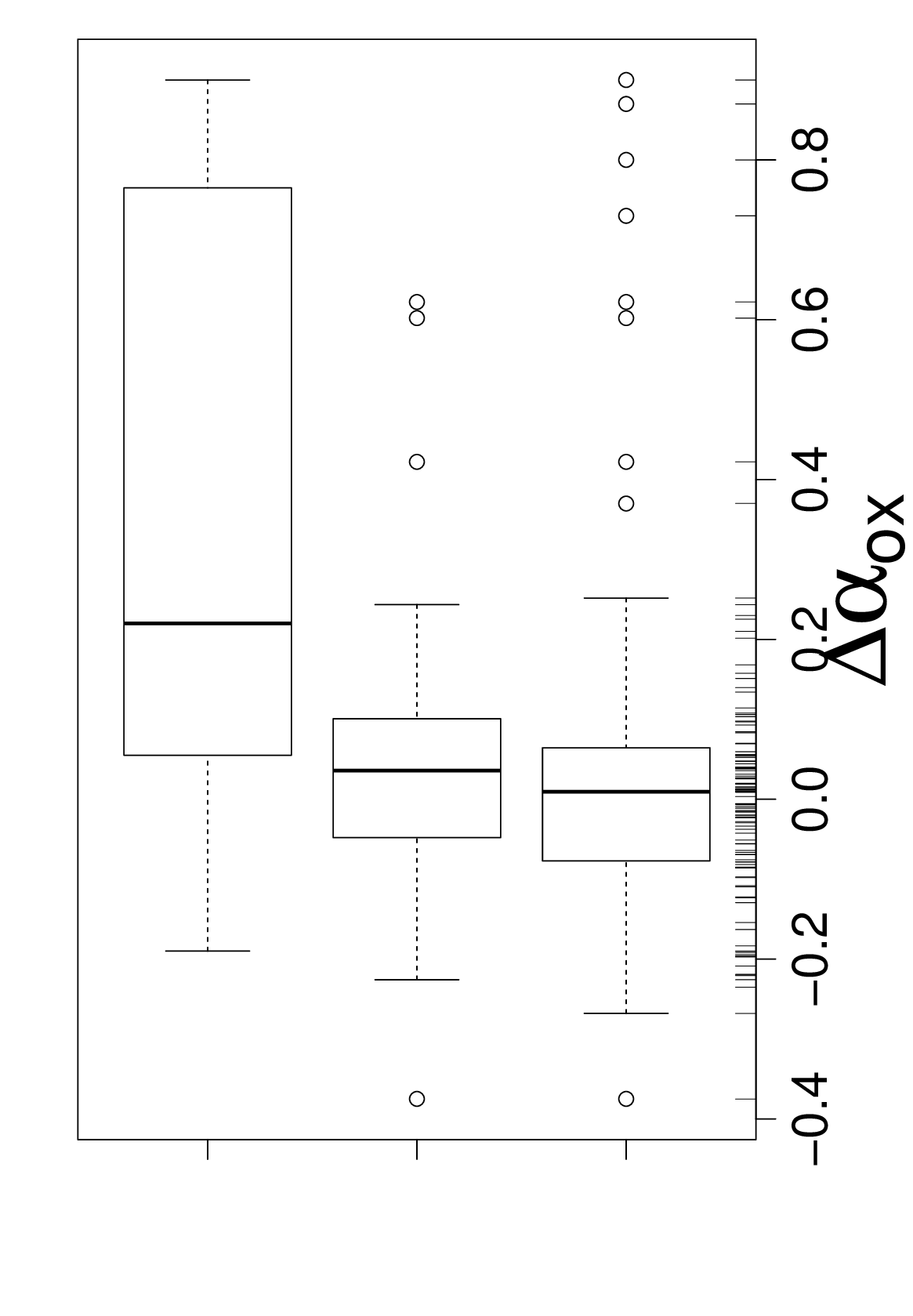}{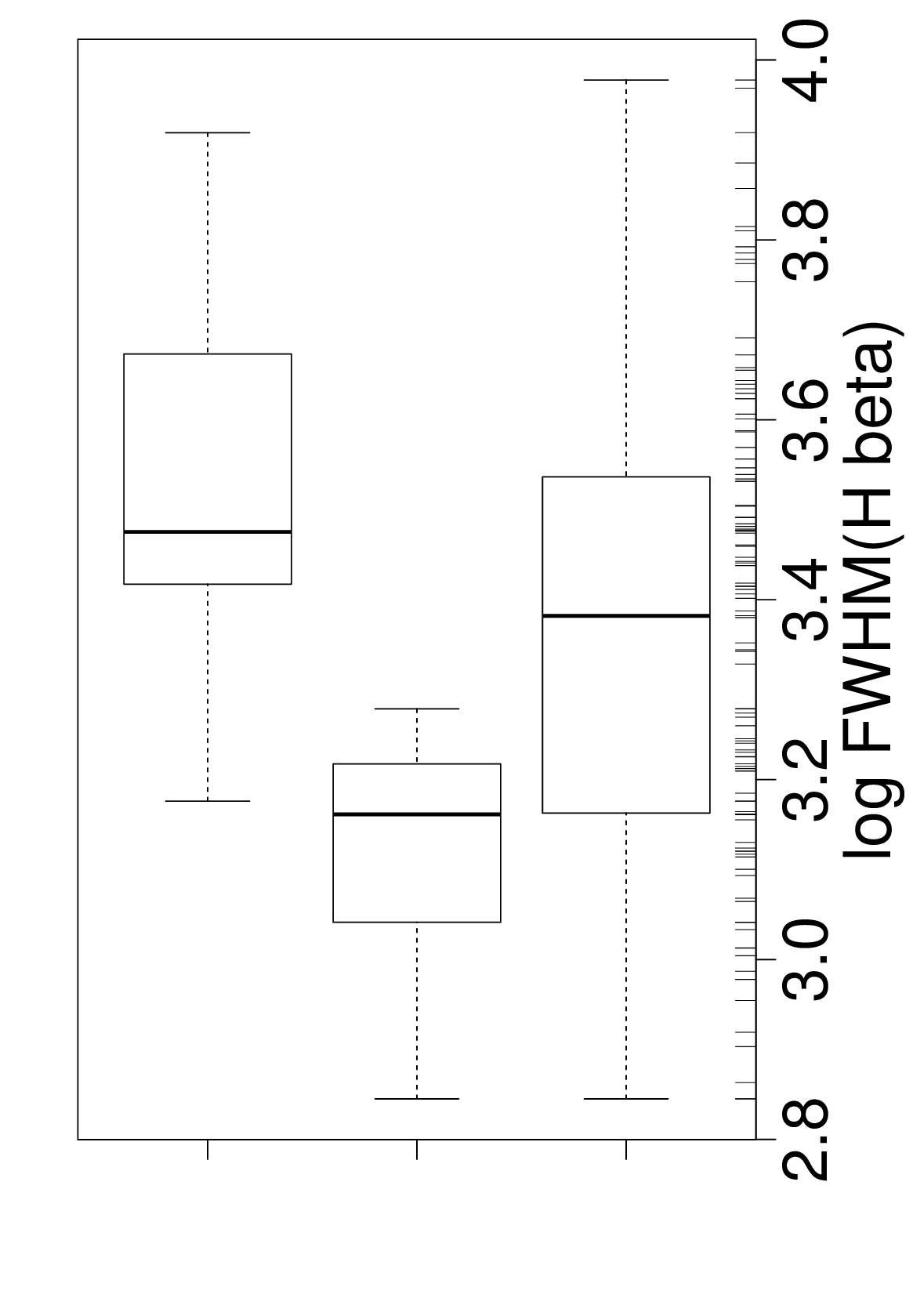}{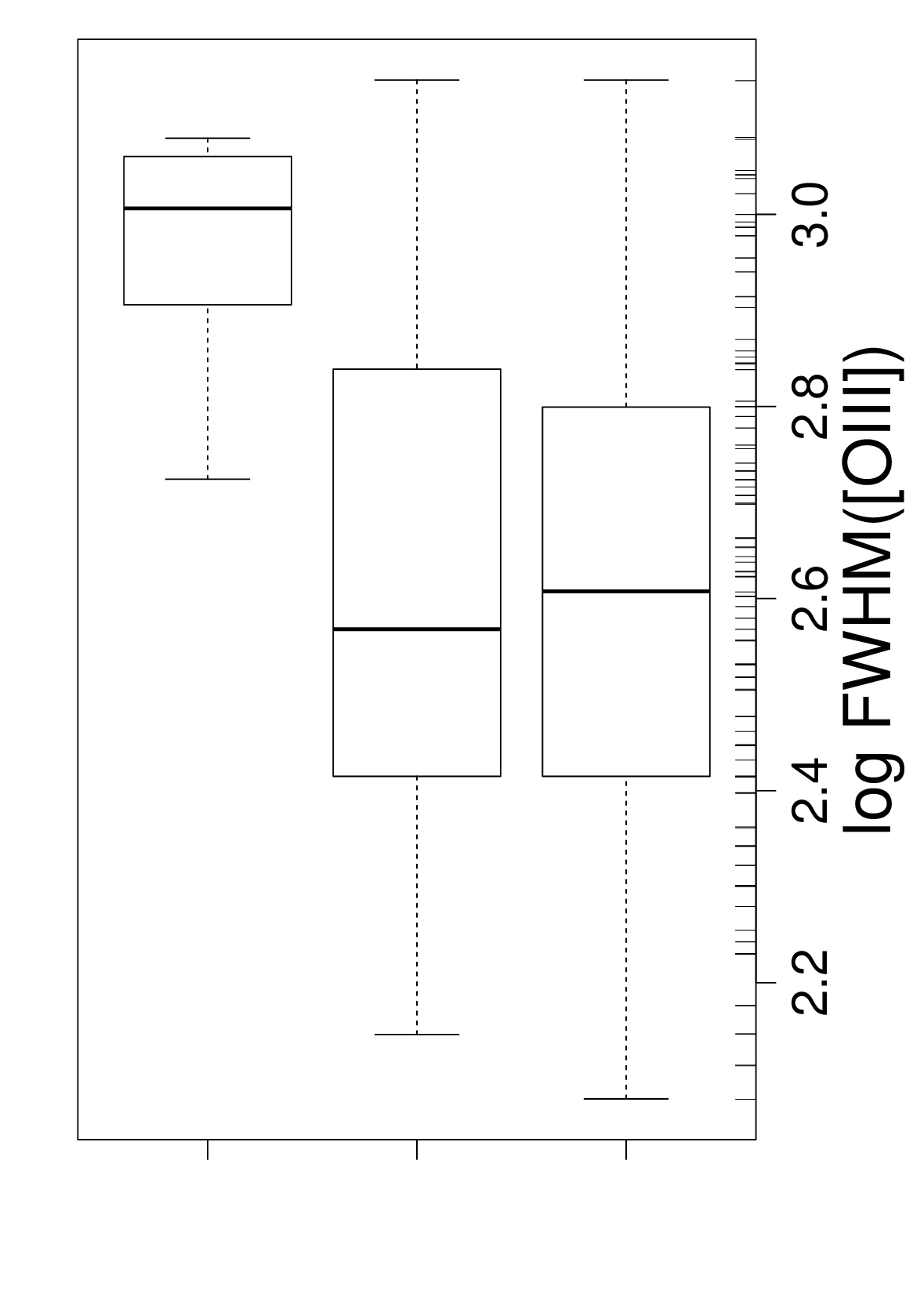}

\plotthreer{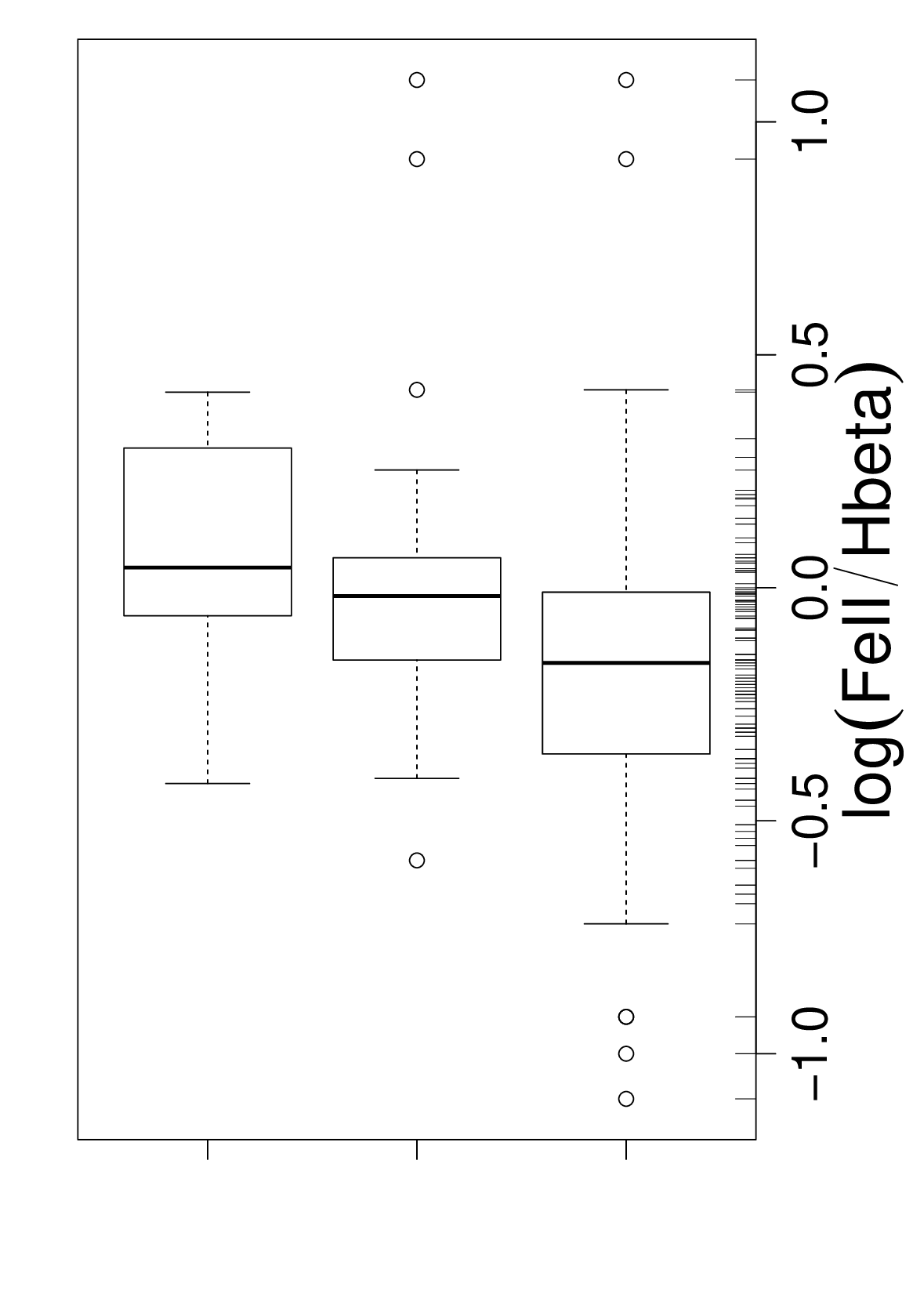}{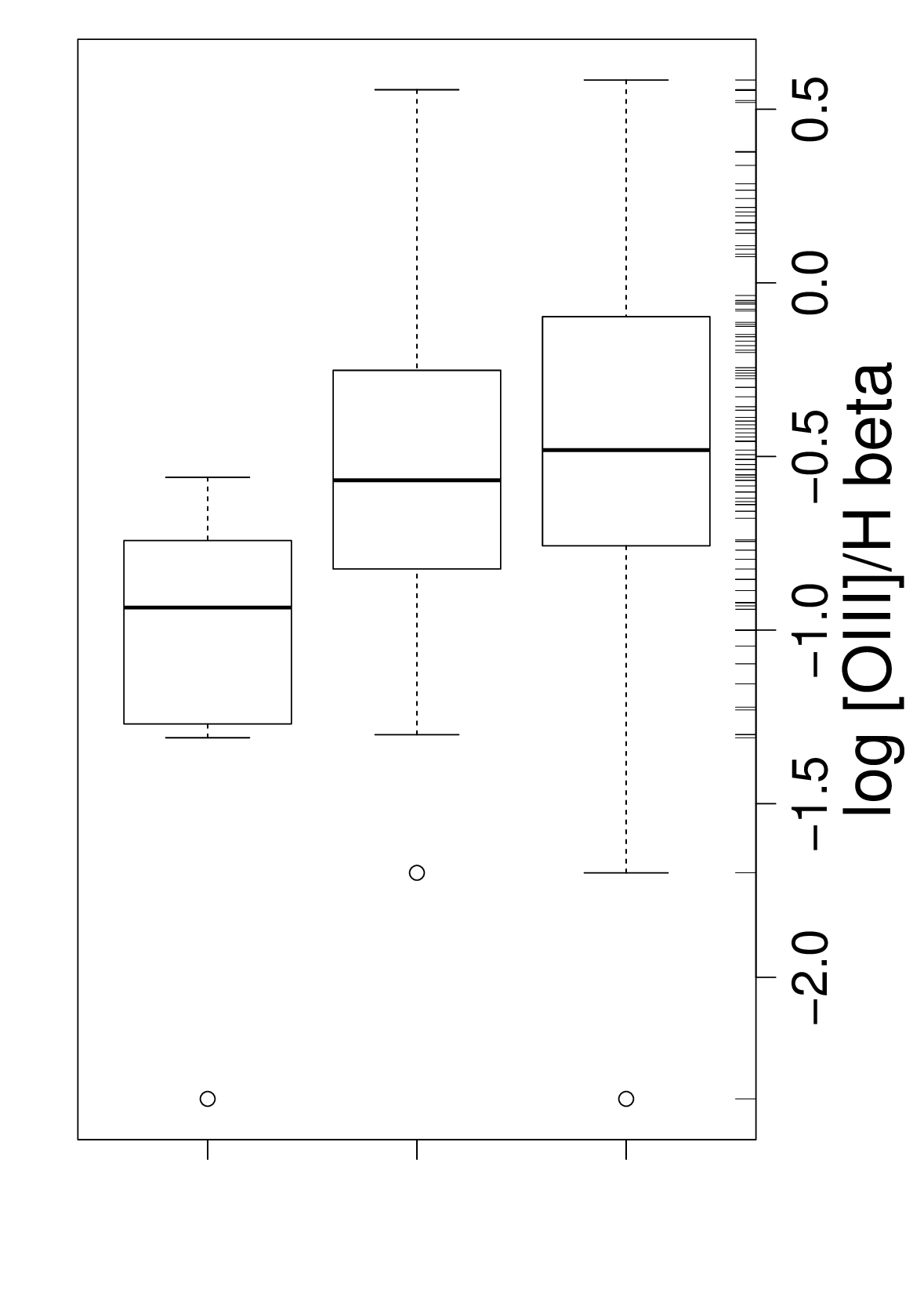}{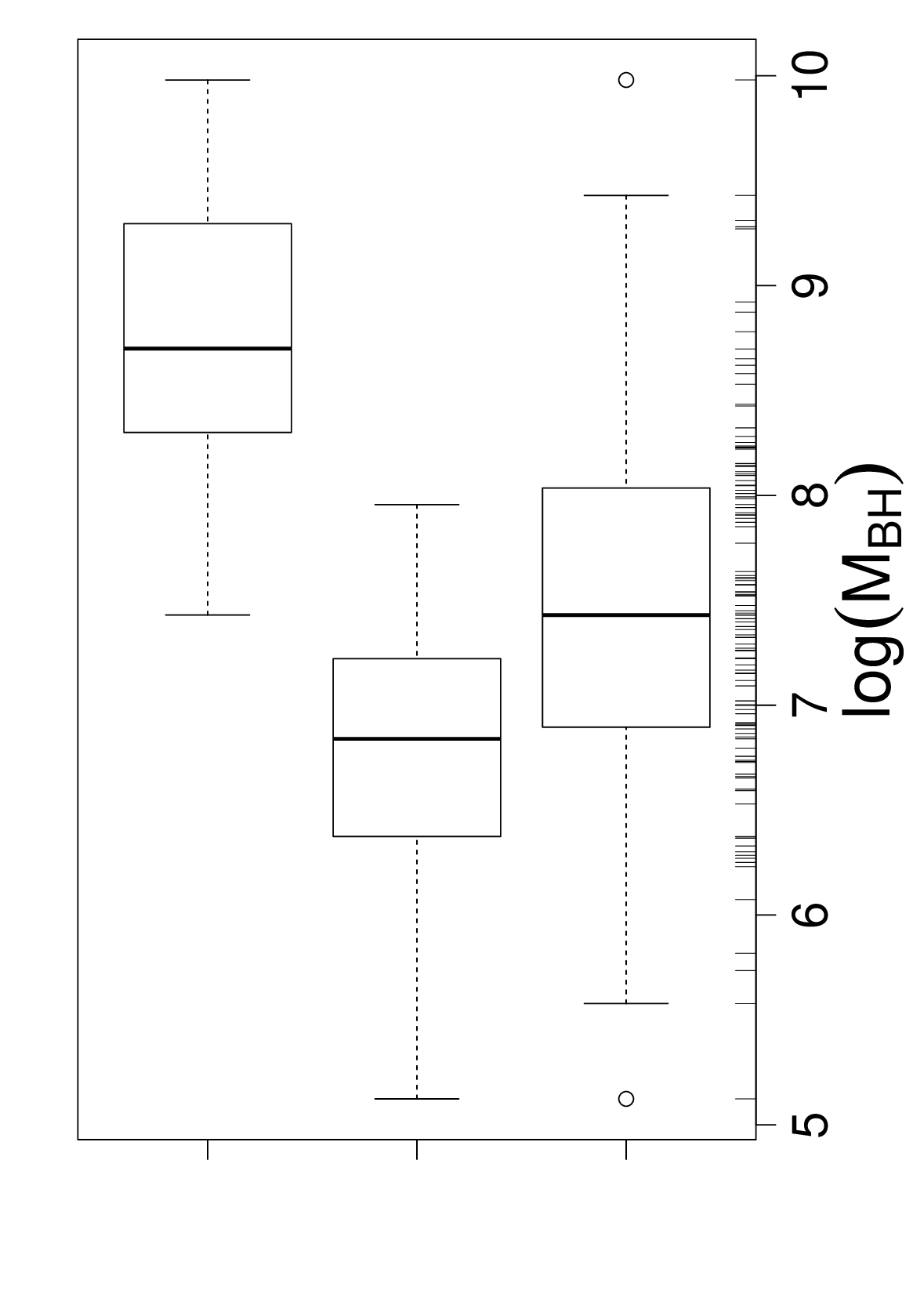}

\plotthreetwo{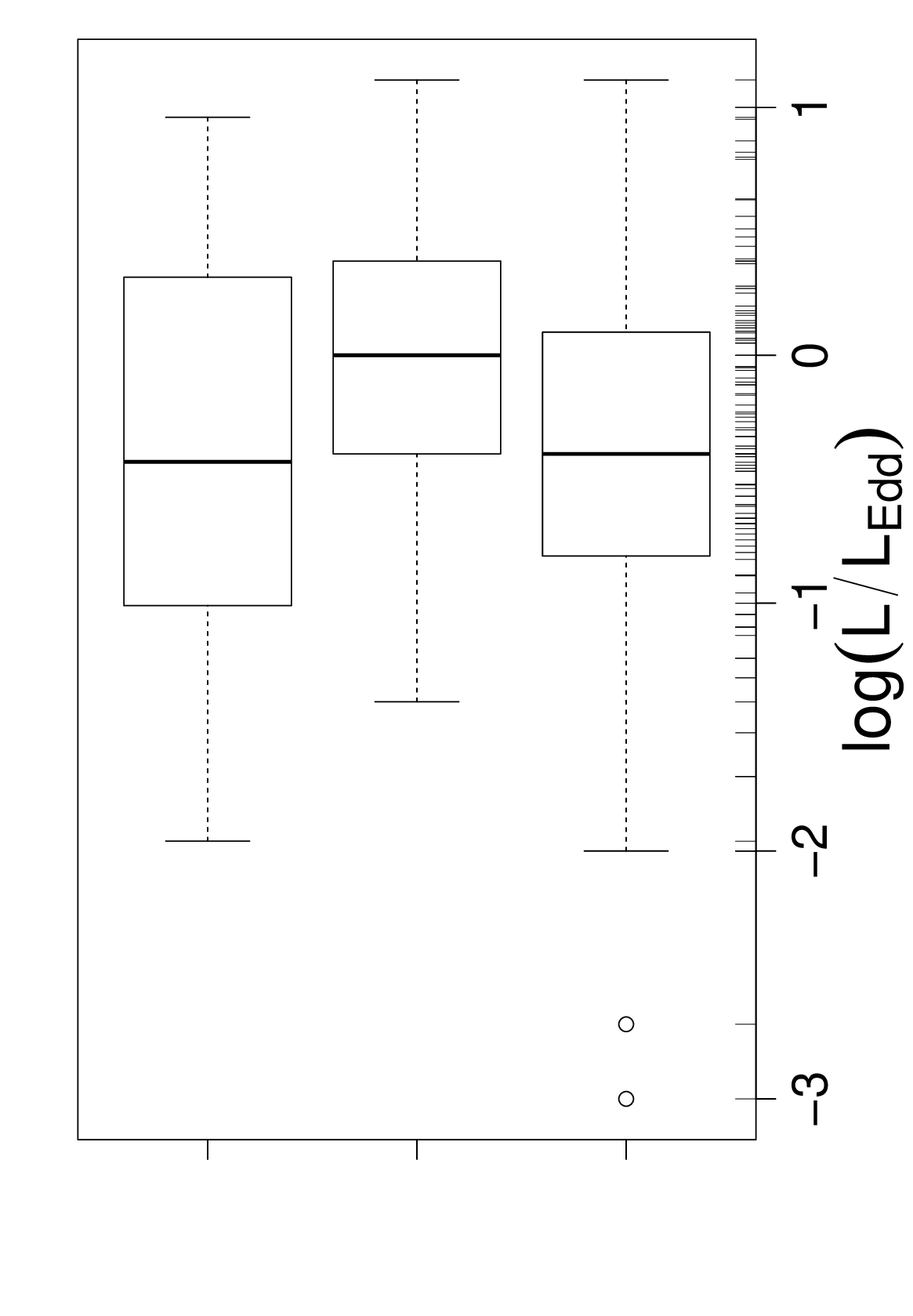}{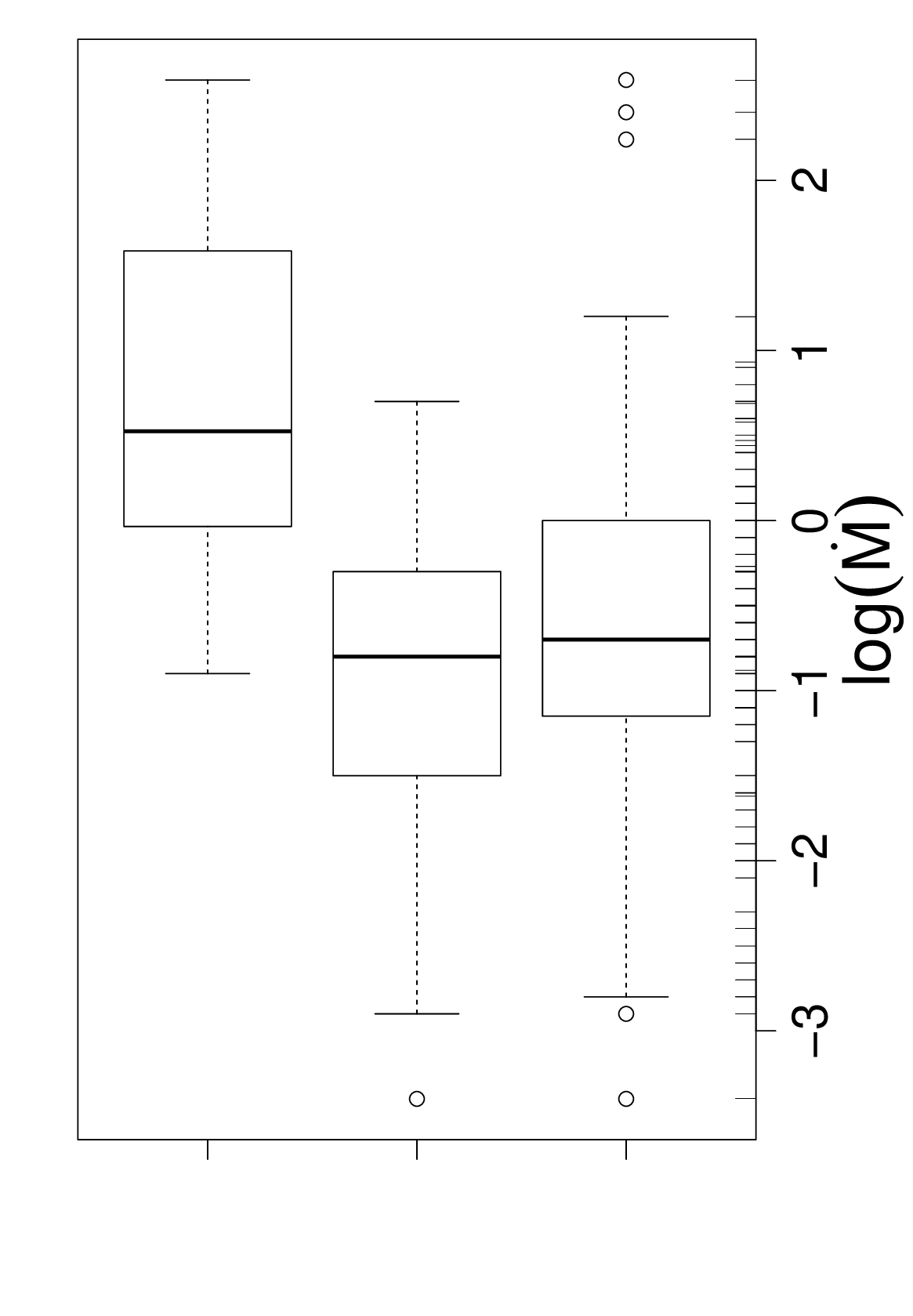}

\caption{\label{plots_distr} Box plots of the distributions 
 of the whole AGN (X-ray selected sample by \citet{grupe10} and
  BAL QSOs) at the bottom, 'classical' NLS1s in the
middle, and BAL QSOs alone at the top. 
}
\end{figure*}

Figure\,\ref{plots_distr} displays the box plots\footnote{
Box plots are a standard visualization tool 
in data mining to obtain information on the distribution of a parameter
\citep[e.g.][]{feigelson12, torgo11, crawley07}.
They allow a simple representation of the parameter distribution and how
 different samples compare. A boxplot consists of three parts: the box, the
 whiskers, and the outliers. 
 The box displays the 1., 2. (median, solid line in the box),  
and 3. quartile of the distribution.
The 'whiskers'  are defined by
the minimum/maximum values of the distribution or the 1.5 times 
the interquartile range (so the width of the box, so basically the 95\%
confidence level),  whatever comes first. Values
beyond the 'whiskers' are outliers and are displayed as circles.}
of the distributions of observed and inferred properties of the 
whole sample (X-ray selected AGN sample of \citet{grupe10} 
plus the \chandra-selected BAL QSOs) on the bottom, 
NLS1s in the middle and BAL QSOs alone at the top. 
The mean, standard distribution, and median of these distributions are summarized in Table\,\ref{agn_statistics}. 

While the UV spectral indices of NLS1s and BAL QSOs are very similar, BAL QSOs show significantly flatter X-ray spectral slopes \ax, and
steeper Optical to X-ray spectral slopes \aox\ than NLS1s. 
There are several ways to interpret these properties. The flatter
   X-ray slopes in BAL QSOs may be a consequence of their slightly lower
   Eddington ratios than NLS1s, as suggested by the relations
   found by \citet{grupe04, grupe10} and \citet{shemmer08}.
      They may also imply stronger (cold or
   ionized) absorption, which was not yet detectable in available
   X-ray spectra due to short exposure times.
The steeper \aox\ can be 
primarily explained by stronger absorption in X-rays in BAL QSOs, 
for example by a partial covering absorber \citep[e.g.][]{grupe03}.
 More importantly, BAL QSOs exhibit much larger $\Delta$\aox\ parameters than NLS1s.

As for the emission line properties, we noticed that the FWHM([OIII]) of BAL QSOs are significantly 
larger than those of NLS1s. Note, that these
broad NLR emission lines are consistent with BAL QSOs being hosted in significantly larger galaxies. 
As we will see later, these [OIII] line
widths are consistent with the black hole masses measured in BAL QSOs.
 The [OIII]/H$\beta$ flux ratios 
in BAL QSOs are significantly lower
than compared with NLS1s.  
The FeII/H$\beta$ ratios of BAL QSOs are larger than those found in NLS1s.  
This follows the 'eigenvector 1' 
relation found by \citet{boroson92}
in which AGN with stronger FeII emission show weaker emission from the NLR. Compared with the BAL QSO
sample of \citet{dietrich09}, the BAL QSOs in our sample here appear quite extreme in their [OIII]/H$\beta$ and
FeII/H$\beta$ line ratios as listed in Table\,\ref{agn_statistics}.
The BAL QSOs in \citet{dietrich09} showed mean, standard deviations and medians of --0.63,
0.29, and --0.65, and --0.12, 0.42, and --0.20 for log [OIII]/H$\beta$ and log FeII/H$\beta$, respectively. These values
are similar to those found for NLS1s (see Table\,\ref{agn_statistics}).
 Note however, that the BAL QSOs in our sample for which optical line ratios could be determined are all at
lower redshift, which the BAL QSOs in the \citet{dietrich09} sample are all at redshifts around z=2.

BAL QSOs have significantly larger black hole masses than NLS1s. 
In general, the black hole masses of BAL QSOs are about 100 times larger than those of NLS1s.
What is somewhat surprising is that the
Eddington ratios \lledd\ of BAL QSOs are about a factor of 10 lower than we  found in the NLS1s in our bright soft X-ray selected AGN sample. 
Figure 7 in \citet{boroson02}, however,
suggested that the Eddington ratios of NLS1s and BAL QSOs are about unity.  

Figure 7 in \citet{boroson02} also suggests that
the mass accretion rates $\dot{M}$ of BAL QSOs and NLS1s are 
significantly different. As displayed in Figure\,\ref{plots_distr} also when
compare our BAL QSO and NLS1 samples we come to a similar conclusion. 
We estimated the  mass accretion rate from the bolometric luminosity with 
$L_{\rm bol} = \eta \dot{M} c^{2}$ assuming 
a mass to radiation efficiency of $\eta$=0.1. 
Also in our sample, BAL QSOs have larger mass accretion rates than NLS1s.
 Note that some of these mass accretion
rates in BAL QSOs exceed 100 \msun/year.

\subsubsection{Correlations}
\begin{figure}
\epsscale{0.75}
\plotone{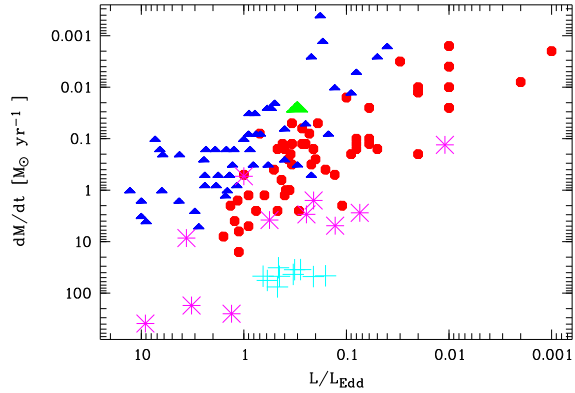}
\caption{\label{plots_lledd_mdot} Eddington ratio \lledd\ vs. mass accretion rate $\dot{M}$. BAL QSOs are displayed as large magenta stars,
NLS1s as blue solid triangles and BLS1s as red solid circles. WPVS 007 is marked as a large solid green triangle. The cyan crosses mark the
 BAL QSOs from the sample of \citet{dietrich09} as a comparison.
}
\end{figure}

Figure\,\ref{plots_lledd_mdot} displays the relation between the Eddington ratio \lledd\ and the mass accretion rate
$\dot{M}$ following Figure 7 in \citet{boroson02}. 
In Figure\,\ref{plots_lledd_mdot} the BAL QSOs of the Chandra sample 
are displayed as magenta stars, NLS1s as solid blue triangles,  BLS1s as
solid red circles, and the BAL QSOs of \citet{dietrich09} are shown as
 turquoise crosses.
Clearly, BAL QSOs and NLS1s fall on two separated groups. 
For a given \lledd, BAL QSOs have a significantly larger mass
accretion rate than NLS1s. This is not completely surprising 
given the fact that the black hole masses of NLS1s is about a 
factor of 100 lower
than that of BAL QSOs (see also the distributions of black hole masses in 
Figure\,\ref{plots_distr}). However, the plot also shows that the
schematic plot shown in Figure 7 in  \citet{boroson02} is too simple. 
Although generally speaking, BAL QSOs have higher mass accretion rates
than NLS1s, as already shown in Figure\,\ref{plots_distr},
 there is some overlap between the two groups and the separation between 
 the two classes of AGN is not as strong as
 suggested in Figure 7 in \citet{boroson02}. 
  The other difference is that
the BAL QSOs in our sample appear to have lower \lledd\ than in 
the \citet{boroson02} sample. 
The lower \lledd\ found in
our BAL QSO sample is similar to those found in the BAL QSOs 
in \citet{dietrich09}.

\subsubsection{Principal Component Analysis \label{pca}}
In order to better understand the relations among the observed parameters we applied a Principal Component Analysis 
\citep[PCA; ][]{pearson1901}.
to the sample
of 119 AGN, including the Swift AGN sample plus the 7 low-redshift BAL QSOs listed in Table\,\ref{chandra_log}. 
A PCA is a standard data mining tool that allows to reduce the number of properties 
that describe a sample of sources from many to a few. 
In a mathematical sense, the PCA  searches for the
eigenvalues and eigenvectors in a correlation coefficient matrix.
A good description for the application of a PCA in astronomy
can be found in \citet{wills99} and \citet{boroson92}.

For our sample we used \ax, \auv, \aox, $\Delta$\aox, 
FWHM(H$\beta$), FWHM([OIII]), [OIII]/H$\beta$, FeII/H$\beta$ as 
input parameters in the PCA.
The results of the PCA are summarized in Table\,\ref{pca_results}. This PCA is somewhat similar 
to the one that we applied to basically the
same AGN sample in \citet{grupe04}, although with slightly different input parameters. 

Figure\,\ref{pca_corr} displays the correlations between eigenvector 1 and \lledd, and
 eigenvector 2 and the black hole mass  and $\dot{M}$. 
 While the eigenvector 1 of our sample 
 can be associated with the Eddington ratio \lledd\ (left panel),
 such as in the PCA performed by \citet{boroson02}, the eigenvector 2 of our sample is clearly dominated by the 
 black hole mass (middle panel). 
  The Spearman rank order correlation coefficients, Student's T-test values, and
 probabilities of a random result of the eigenvector 1 - \lledd\ and eigenvector
 2 - $M_{\rm BH}$ correlations 
 are $r_{\rm s}=-0.52$, $T_{\rm s}=-6.62$, $P<10^{-8}$, and 
  $r_{\rm s}=-0.82$, $T_{\rm s}=-15.04$, $P<10^{-8}$, respectively.
  
 Although the
 mass accretion rate is anti-correlated with eigenvector 2 (right panel) with 
 $r_{\rm s}=-0.32$, $T_{\rm s}=-3.64$, $P=4.1\times 10^{-4}$,
 it is in our sample also anti-correlated
 with eigenvector 1 ($r_{\rm s}=-0.44$, $T_{\rm s}=-5.32$, $P=5.2\times 10^{-7}$
  However, eigenvectors are orthogonal, 
  which means they should not show
 a correlation among them. 
 Therefore we conclude that in our
 sample eigenvector 2 represents the black hole mass and not $\dot{M}$.
 Note, that for his input sources and input parameters, \citet{boroson02} concluded from his PCA that
 eigenvector 2 represents the mass accretion rate $\dot{M}$.
 As a test we performed a PCA with just the optical line properties (FWHM(H$\beta$), FWHM([OIII]),
 [OIII]/H$\beta$, and FeII/H$\beta$). In this case the correlation between eigenvector 2 and $\dot{M}$ becomes
 more significant with $r_{\rm s}=-0.46$, $T_{\rm s}=-5.67$, $P=1.0 \times 10^{-7}$. However, the correlation between
 eigenvector 2 and the black hole mass is still stronger than that with $\dot{M}$.

\section{\label{discuss} Discussion}

The main motivation behind this study is the question of whether BAL QSOs and
NLS1s are intrinsically similar 
but only appear at different ends of the black hole mass
distribution spectrum as suggested by \citet{brandt00b} and \citet{boroson02}. 
Although we see similarities between our sample and that of \citet{boroson02}, 
in particular that NLS1s and BLA QSOs have similar FeII/H$\beta$ and [OIII]/H$\beta$ line ratios
we also notice differences in the studies.

\subsection{Selection Effects}
While our sample size is significantly larger than the initial exploratory
  study of \citet{boroson02}, it is still subject to selection effects, and
  we therefore first address the key question, if and how our sample selection
  affects the parameters we wish to compare among NLS1s and
  BAL-QSOs:
  
\begin{enumerate}
\item {\bf Redshift Distributions:}
Traditionally, BAL QSOs have been discovered at redshifts of about z=2 \citep[e.g.][]{weymann91}.
This leads to
selecting highly-luminous and therefore high black hole mass AGN. On the other hand, our AGN sample consists of bright
X-ray selected AGN which appear at much lower redshifts with z$<$0.4 \citep{grupe04a, grupe10}. Nevertheless, the 7 low
redshift BAL QSOs presented in this paper have redshifts that are between z=0.04 and z=0.47 and fall into 
the redshift range of our AGN sample. 
This redshift range is also comparable to the PG sample by \citet{boroson92}
which contains the BAL QSOs used in \citet{boroson02} sample. The 7 low-redhshift BAL QSOs in our sample allow a
direct comparison to the low-redshift AGN sample by \citet{grupe04a} 
and \citet{grupe10} which is not necessarily
true for the high-redshift objects.

\item {\bf Soft X-ray selection of the AGN sample:} Our AGN sample was selected by their X-ray properties: X-ray bright
and soft X-ray spectra \citep{grupe98, grupe01, grupe04a, grupe10}. A steep X-ray spectrum, however, is a consequence of
high Eddington ratios \citep{grupe04, grupe10, shemmer08}. 
Therefore the soft X-ray selection favors sources with
  high L/Ledd among the class of NLS1 galaxies as a whole and 
the Eddington ratios found in our AGN sample may
appear larger than in an optically selected NLS1 sample.
 The soft X-ray selection also misses highly absorbed AGN, such as typically BAL QSOs. 

\end{enumerate}

Although we have to keep these selection effects in mind, they do not make our study invalid. 
On the contrary, our
comparison is between BAL QSOs and NLS1s, and the latter are usually found among soft X-ray selected AGN. It has
also been shown that in BAL QSOs and in NLS1s absorbers can develop over time and the absorption column density and
covering fraction can vary in X-rays as well as in the UV. 
For NLS1s a good example here is WPVS 007 which has developed strong BALs within a decade
\citep{leighly09}. On the other hand, more and more evidence has been found that also the BALs in BAL QSOs can be
highly variable \citep[e.g.]{filiz12, filiz13, capellupo11, capellupo12}.

\subsection{Mass accretion rate $\dot{M}$}
 While Figure 7 in \citet{boroson02} 
suggests that NLS1s and BAL QSOs 
are both at high \lledd\, where BAL QSOs have significantly larger mass accretion rates than NLS1s,
 we found from our sample that this picture
is more complicated:
 Although also in our sample BAL QSOs appear to show the highest mass accretion rates, some appear to
  be significantly lower, even down to 0.1
 \msun/year in the case of PG 1004+130. Compared with the BAL QSOs in \citet{dietrich09}, where the mean mass 
 accretion rate was of
 the order of about 35 \msun/year, the mass accretion rates in the sample presented here 
 appear to be rather modest with a median of about 4 \msun/year. However, some of the mass accretion rates of the
 BAL QSOs in our sample are extreme, exceeding 100 \msun/year. 
The distribution of the mass accretion rates in BAL QSOs spans over three orders of
magnitude as shown in Figure\,\ref{plots_distr}. 
Although the mass accretion rate of NLS1s also stretches over three orders of
magnitude, the lowest values of the mass accretion rate are of the order of
10$^{-3}$ solar masses and the highest of the order of
 a few solar masses per year. There is a
some overlap between NLS1s and BAL QSOs in mass accretion rate in our sample.
The two samples are not as separated as the BAL QSO and NLS1s samples of
\citet{boroson02}.
This maybe in part a selection effect, because the NLS1s is our sample are soft X-ray selected 
which are primarily sources with high \lledd.

If we just focus on the 7 low-redshift BAL QSOs in our study (see Table\,\ref{chandra_log}) we notice that the mass
accretion rates of these BAL QSOs are in the
 range between 0.1 to 8 \msun/year and are somewhat comparable to the mass
accretion rates seen in NLS1s. The BAL QSOs which really make a difference are those at higher redshifts.

 We noticed
 that the mass accretion rates in BAL QSOs is rather high
with more than 100 $M_{\rm sun}$ yr$^{-1}$, and 35 in the BAL QSO sample by \citet{dietrich09}.
One reason why these 
accretion rates appear so high might be our assumption of the mass to radiation
efficiency which we for simplicity just set to $\eta=0.1$. 
If however the black hole spin in BAL QSOs is significantly higher this would increase the
efficiency and reduce the mass accretion rate necessary to explain the observed bolometric luminosity. 
For example, if the black hole in a BAL
QSO is maximal spinning (a=0.998) the efficiency will be $\eta=0.32$, 
or three times higher than our assumed $\eta$=0.1.  
This would imply that
the mass accretion rate is a factor of 3 lower, 
which is much easier to explain than the 
high mass accretion rates we estimated at first. 
In contrast to BAL QSOs, in NLS1s the
efficiency may be lower and the mass accretion rate higher if the black hole
 is rotating slower than in our assumption. 
Keep in mind that for an efficiency of $\eta=0.1$ the black hole is 
spinning already with a=0.6. 
One argument that the
black hole spins in BAL QSOs and NLS1s are systematically
 different, is the finding that the black hole masses 
in BAL QSOs are significantly
larger than in NLS1s. A larger black hole mass also means that the accreted matter has transferred a larger
 amount of angular momentum to this
black hole which then results in a larger spin (in a scenario where SMBHs primarily grow
    by accretion).
One more argument for a higher efficiency of the 
mass to radiation conversion due
to a relativistically spinning black hole is that the mass accretion rates in the sample presented here appear to 
be relatively low compared with
those of the BAL QSO sample by \citet{dietrich09}.

\subsection{Eddington ratios \lledd}
The BAL QSOs in our sample also have Eddington ratios \lledd\ which are  lower than those of NLS1s.
 In Figure 7 in
\citet{boroson02} shows that the BAL QSOs in the \citet{boroson92}
sample are all at \lledd=1, which is typical for NLS1s.
Although our AGN/NLS1 sample is soft X-ray selected and has therefore a bias towards high \lledd\ AGN, the Eddington
ratios found in our sample agree well with those of the NLS1s in the optically selected sample by \citet{boroson92}.
However, a comparison
 between our sample and that of
\citet{boroson02} appears somewhat difficult because 
\citet{boroson92} do not list the 
values of \lledd\ and $\dot{M}$ of the NLS1s and BAL QSOs 
of their PG quasar sample.

Although the eigenvector 1 in our PCA
can also be associated with the Eddington ratio \lledd\ as 
in the \citet{boroson02} sample, our eigenvector 2 seems to be
different. While \citet{boroson02} suggested that the eigenvector 2 in his sample is clearly
 associated with the mass accretion rate $\dot{M}$, in our
sample eigenvector 2 is more
strongly correlated with the black hole mass. These 
differences  are not surprising. A PCA is a purely
mathematical tool and the results depend strongly on the set of
input parameters as well as the objects in the sample. 
While in our PCA we used continuum properties and line widths
and ratios, the PCA in \citet{boroson02} primarily used optical line properties. 
Therefore it can be expected that the results 
 of the PCAs could be different.
However, when we only use optical line parameters in a PCA, we still find that in our sample that eigenvector 2 is
strongly correlated with the black hole mass.

\subsection{Black Hole Masses and Growth}

From our BAL QSO and NLS1 samples we found that BAL QSOs have black hole masses more than 100 times larger than 
those seen in a typical NLS1. At first glance this seems to be a  selection effect because most BAL QSOs appear
at higher redshifts which leads to selecting objects with higher 
luminosities and
larger back hole masses. However,  for low-redshift
 BAL QSOs in our
\chandra\ sample, we note that even these objects have black hole 
masses of the order of $10^9$\msun, 
significantly larger than those found in NLS1s (see Figure\,\ref{plots_distr}). 

In contrast to NLS1s, 
BAL QSOs may have already fully
developed black holes with regards to their host galaxy masses. This argument is
supported by the M - $\sigma$ relation which relates the mass of the central
black hole to that of the host galaxy. BAL QSOs
fall right on the \citet{tremaine02} M - $\sigma$  relation as shown in  
in Figure\,\ref{plots_m_sigma} (the \citet{tremaine02} 
relation is shown as a black solid line), while NLS1s typically fall below this
relation \citep{grupe04b}. 
We adopted the method introduced by
\citet{nelson00} who suggested to use the [OIII] line as a proxy of the stellar velocity 
dispersion in the host galaxy bulge\footnote{We assign the expression $\sigma_{\rm [OIII]}$ for the stellar velocity
dispersion determined from the width of the [OIII]$\lambda5007\AA$ line.}.
Because the
Narrow Line Region is in  the gravitational potential of the host galaxy bulge, the NLR gas follows
 the same velocity dispersion as the bulge stars.
As a comparison we also show the BAL QSOs from the study by \citet{dietrich09} in this plot. 
Note that the BAL QSO of the \citet{dietrich09}
sample have redshifts between z=1-2 and from our BAL QSO sample we have only [OIII] and H$\beta$ measurements 
of the 7 low-redshift BAL QSOs.
The BAL QSOs of our sample which are at higher redshifts also have similar black hole masses as the 
objects in the \citet{dietrich09} BAL QSO
sample. Whether NLS1 galaxies, as a group, fall on or below the
 $M-\sigma_{\rm [OIII]}$ relation is an important question
 \citep[e.g.][]{grupe04b,mathur05a, mathur05b, komossa07} 
The only BAL QSO that falls below the M - $\sigma_{\rm [OIII]}$ relation is PG 1001+054. 
The (soft X-ray selected) NLS1 sample of \citet{grupe04b} was located
    significantly below the $M-\sigma_{\rm [OIII]}$ relation, while the NLS1 sample of
    \citet{komossa07} was consistent with the $M-\sigma_{\rm [OIII]}$ relation of non-active galaxies,
    with the exception of sources at high \lledd. These had their whole [OIII] emission
    systematically blueshifted (``blue outliers''), implying large outflows which come with
    a strong extra line broadening apparently displacing them from the $M-\sigma_{\rm [OIII]}$ relation.
Therefore the FWHM([OIII]) measured in these objects may not represent the velocity of the velocity 
dispersion in the bulge, but has an
additional, unknown contribution by outflows. In the same way as turbulent motion can lead to an 
overestimation of the the black hole masses from
BLR emission line widths as shown by \citet{kollatschny11}, an outflow adds an additional velocity 
component to the total velocity and therefore
leads to an overestimation of the mass of the host galaxy bulge. 

\subsection{Outflows in BAL QSOs and NLS1s}
The main question remaining is why do we see strong outflows which appear as deep UV absorption troughs 
in BAL QSOs but these are rarely seen in NLS1s. In a matter of fact, the only NLS1s that has shown dramatic 
UV absorption roughs is WPVS 007. Part of the answer maybe that BAL QSOs are 
highly-luminous AGN and as 
shown by \citet{laor02}, the strengths of the BAL troughs correlates with luminosity. Nevertheless,
the NLS1
 WPVS 007 is a low 
luminosity AGN and seems to contradict these findings. Strong outflows are typically associated with 
high \lledd\ which results in a strong radiation field from the accretion disk producing radiation driven winds/outflows
\citet[e.g.][]{proga04, kurasawa09a,kurasawa09b}. On the other hand, NLS1s with high \lledd\ exhibit strong outflows
through their broad, blueshifted [OIII] emission lines \citep[e.g.][]{bian05, zhang11, xu12}, which we do not see in 
BAL QSOs \citep[e.g.][]{dietrich09}.

\subsection{Overall Comparison and Conclusions} 
 
One reason for the slight 
differences between our study and that by \citet{boroson02} maybe the different
sample sizes. While in our sample we have 11 BAL QSOs (7 at low redshift) and 53
NLS1s, the numbers in the \citet{boroson02} sample are 4 BAL QSOs and 8 NLS1s. The low number of in particular
BAL QSOs may 
have a large influence on these differences.
 As mentioned about, the mass accretion rates in other BAL QSO samples, 
like that in \citet{dietrich09} appear to be larger.
Therefore in a sample of just 4 sources, the chance of selecting BAL QSOs with larger mass
 accretion rates  is quite high.

To conclude, we confirm the results by
\citet{brandt00b} and \citet{boroson02}  about the differences in the black hole mass 
distributions between BAL QSOs and NLS1s. We also confirm 
 that the Eddington ratios of BAL QSOs and NLS1s are at
the higher end of the \lledd\ distribution of AGN in general. 
However, from our samples of NLS1s and BAL QSOs the Eddington ratios
NLS1s typically operate around unity, while those of the BAL QSOs are slightly lower.  
NLS1s and BAL QSO also share that their FeII/H$\beta$ ratios are very high and
their [OIII]/H$\beta$ ratios are low - following the 'classic \citet{boroson92}' eigenvector 1 relation. 
Although some BAL QSOs in our sample exhibit very high mass accretion rates inferred from their bolometric luminosities,
the mass accretion rates of BAL QSOs and NLS1s appear to be more similar than previously thought. This is in particular
true when we look at the low redshift BAL QSOs in our sample.  
 Interestingly, even though BAL QSOs have relatively large \lledd\ ratios, they
fall right onto the $M_{\rm BH} - \sigma_{\rm [OIII]}$ relation suggesting that they have fully developed black holes. 
One reason for the large bolometric luminosities found in BAL QSOs 
may be that the black hole is spinning
relativistically which leads to a significantly higher efficiency of the mass to radiation conversion.  
In the future we need to increase the number of BAL QSOs in our 
study to obtain better statistics on the relations found in our study. In
particular we need to obtain more X-ray observations of BAL QSOs 
to measure the continuum properties. And last but not least, in the far future
we need to study the Fe K$\alpha$ line emission line to study the gravitational and relativistic 
effects which can be used to measure the black hole spin. 

\acknowledgments

We would like to thank Neil Gehrels for approving our ToO requests and
the \swift\ team for performing the ToO observations. We thank our anonymous referee for useful comments and
suggestions which have improved the paper significantly. 
In particular we want to thank S. Komossa for very valuable discussions and comments on this paper.
Many thanks to Bev Wills 
for providing the optical spectrum of IRAS 07598+6508
\citep{hines95}. We also thank Matthias Dietrich for providing the optical line
fluxes for the BAL QSOs reported in \citet{dietrich09}, and Leisa Townsley for 
supporting this project.
This research has made use of the NASA/IPAC Extragalactic
Database (NED) which is operated by JPL,
Caltech, under contract with NASA. This publication makes use of data products from the 
Wide-field Infrared Survey Explorer, which is a joint project of the University of 
California, Los Angeles, and JPL, 
funded by NASA. 
This publication makes use of data products from the Two Micron All Sky Survey, which is
 a joint project of the University of Massachusetts and the Infrared Processing and Analysis
  Center/California Institute of Technology, funded by NASA and NSF.
  The scientific results reported in this article are based partly on  
  data obtained from the Chandra Data Archive.
  This research has made use of software provided by 
  the Chandra X-ray Center (CXC) in the application package CIAO.
This research has made use of the
  XRT Data Analysis Software (XRTDAS) developed under the responsibility
  of the ASI Science Data Center (ASDC), Italy.
\swift\ at PSU is supported by NASA contract NAS5-00136 (DG + JN).
This work was supported by the Penn State ACIS Instrument Team Contract SV4-74018, issued by the Chandra X-ray 
Center, which is operated by the Smithsonian Astrophysical Observatory for and on behalf of NASA under 
contract NAS8-03060. The Guaranteed Time Observations (GTO) included here were selected by the ACIS
 Instrument Principal Investigator, Gordon P. Garmire, of the Huntingdon Institute for X-ray Astronomy, 
 LLC, which is under contract to the Smithsonian Astrophysical Observatory; Contract SV2-82024.

\clearpage

\begin{figure}
\epsscale{0.75}
\plotone{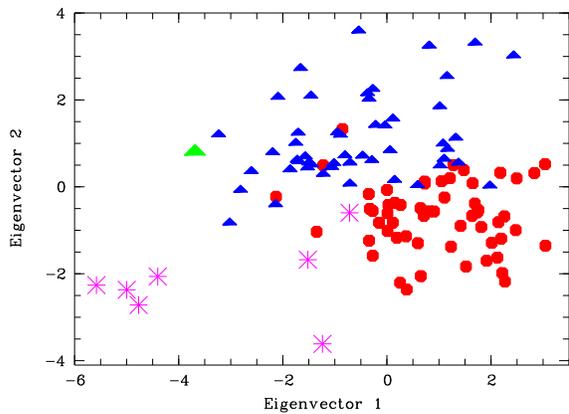}
\caption{\label{plots_ev1_ev2} Eigenvector 1 vs. Eigenvector 2 diagram. The symbols are the same as defined in Figure\,\ref{plots_lledd_mdot}.
 }
\end{figure}

\begin{figure*}
%\epsscale{0.75}
\epsscale{1.5}
\plotthree{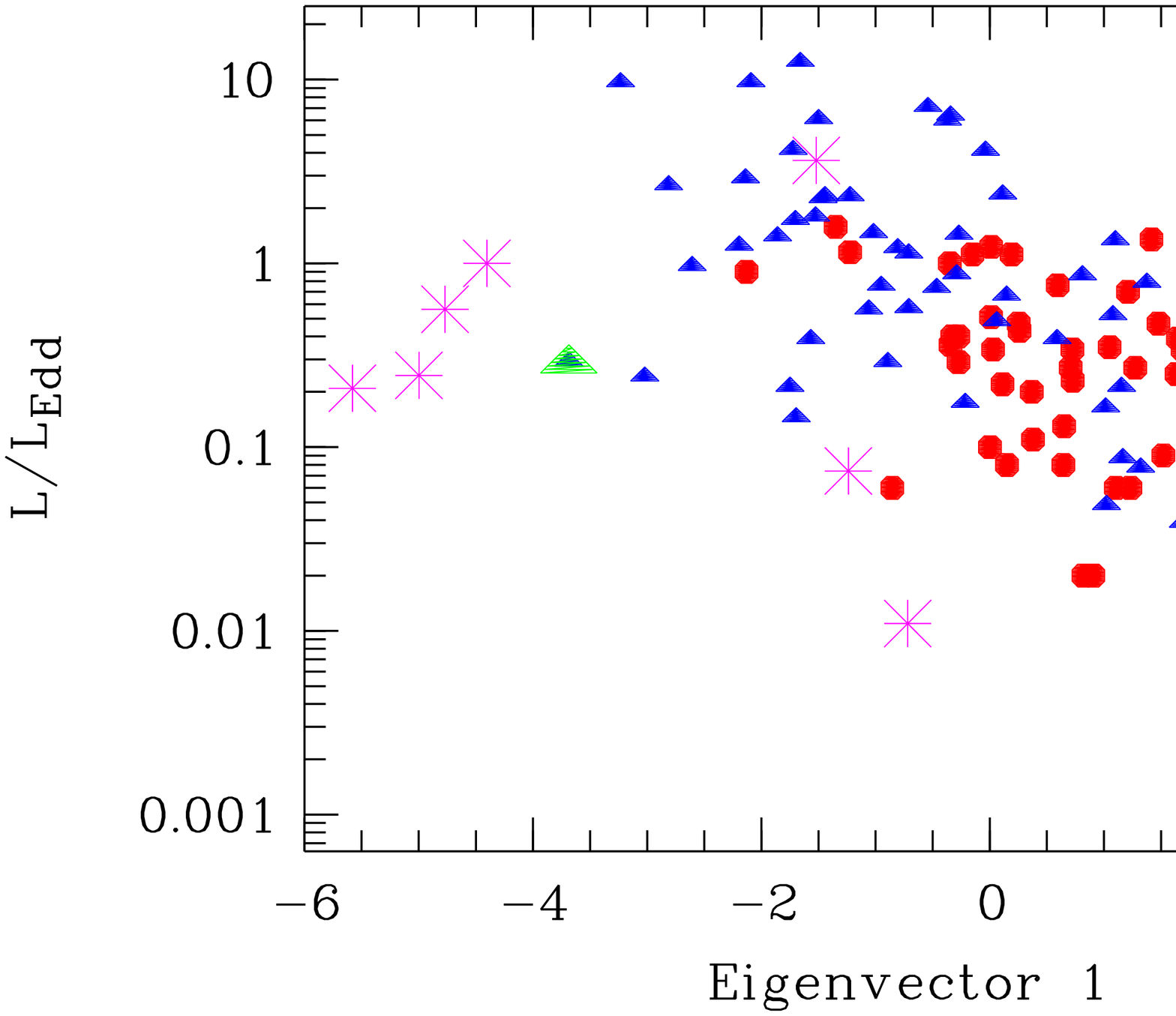}{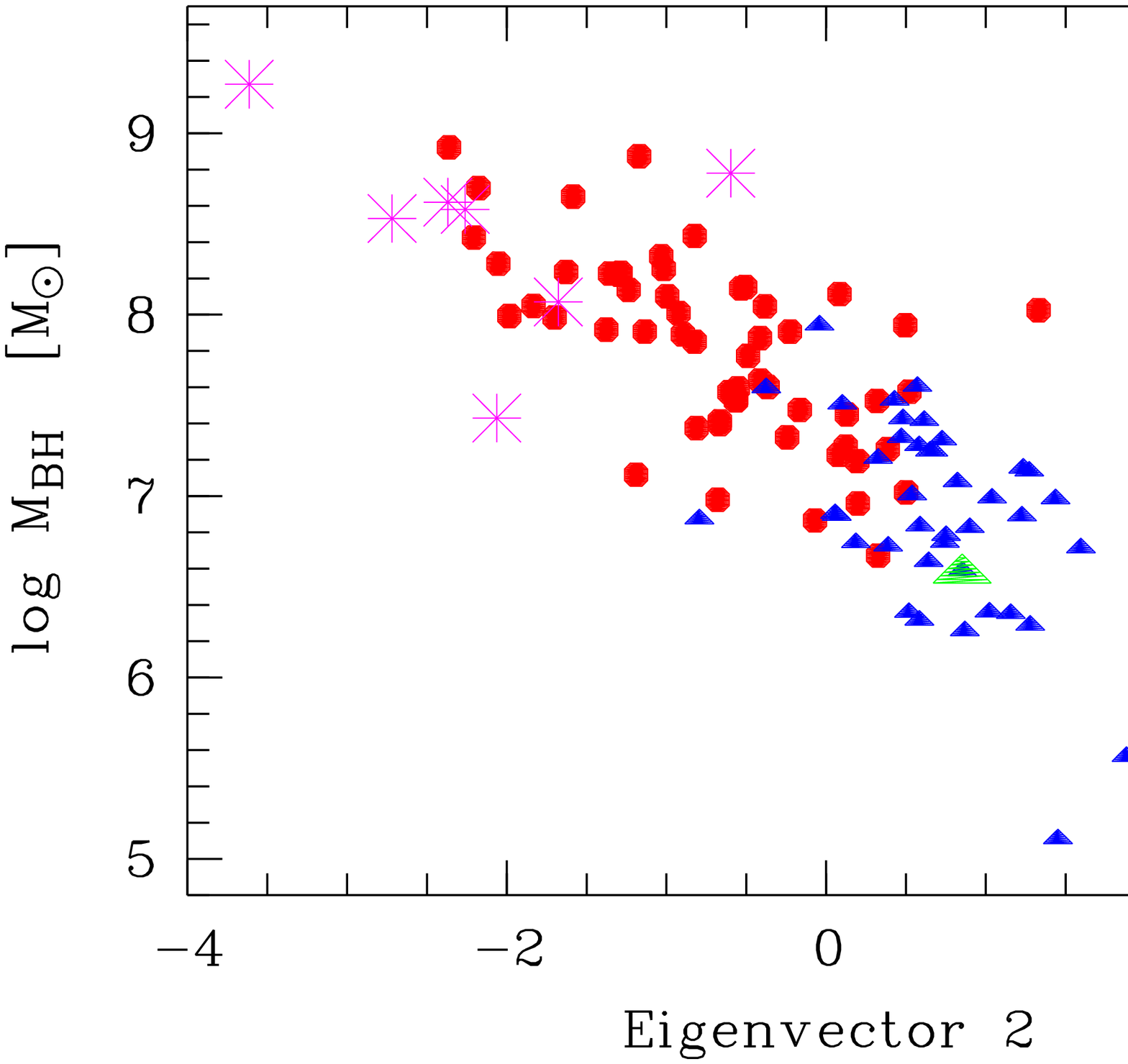}{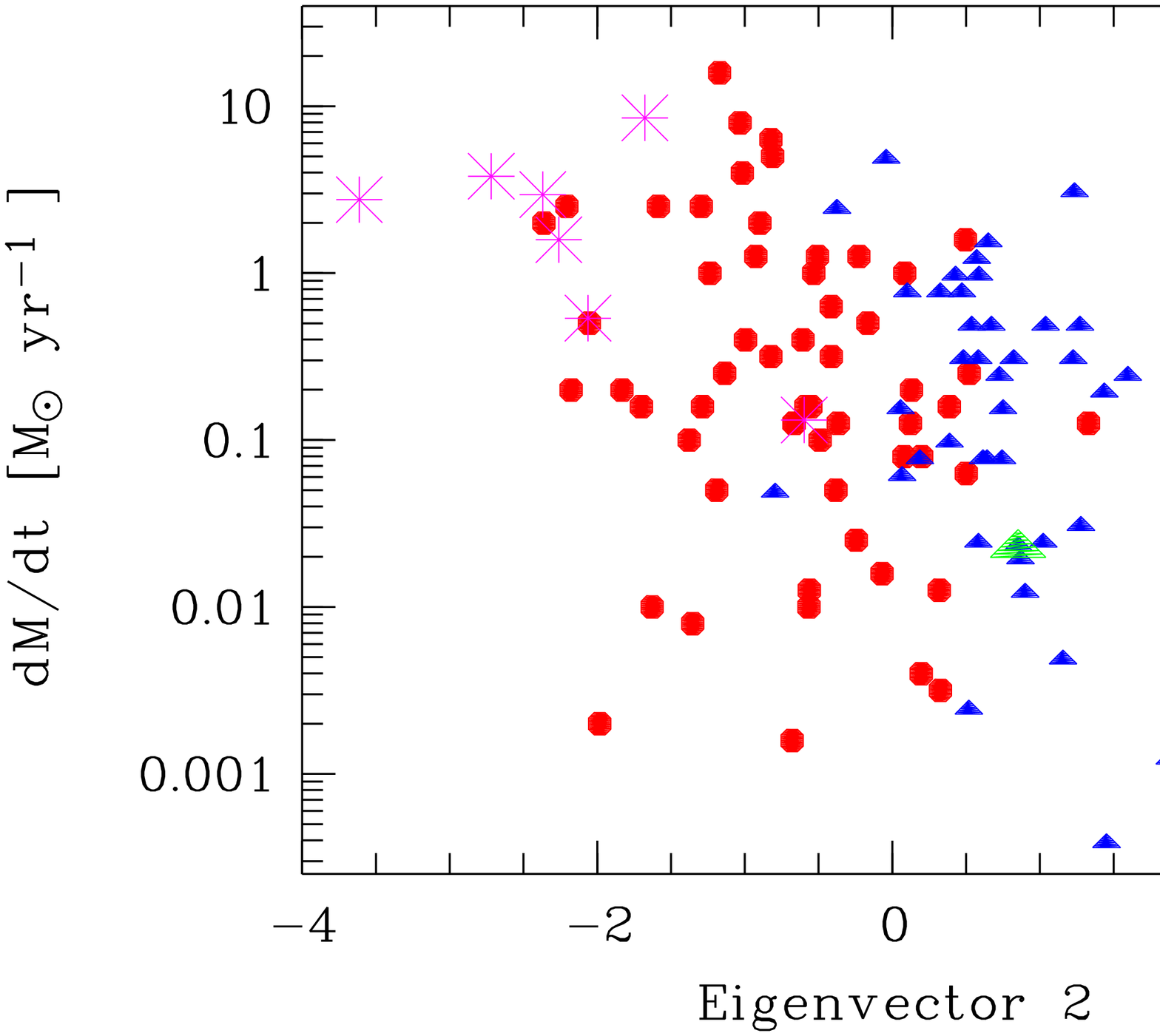}

\caption{\label{pca_corr} Correlations between Eigenvector 1 and \lledd, and Eigenvector 2 and black hole mass $M_{\rm BH}$ and the mass
accretion rate $\dot{M}$.
}
\end{figure*}

\begin{figure}
\epsscale{0.75}
\plotone{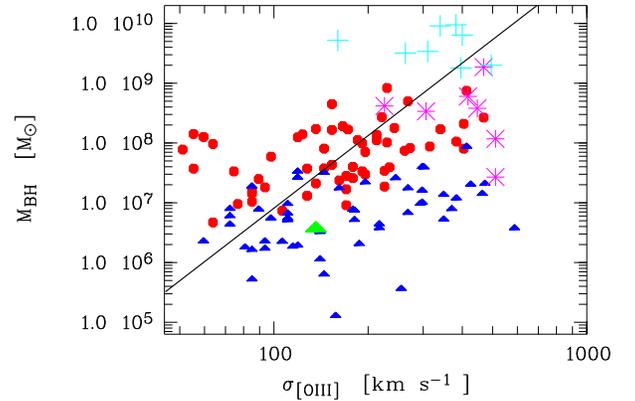}
\caption{\label{plots_m_sigma} M - $\sigma$ relation for the AGN and BAL QSOs in the sample. 
 The symbols are the same as defined in Figure\,\ref{plots_lledd_mdot}. The cyan crosses mark the
 BAL QSOs from the sample of \citet{dietrich09} as a comparison.
 }
\end{figure}

\clearpage

\begin{deluxetable}{lcccrccccccr}
\rotate
\tabletypesize{\tiny}
\tablecaption{Object summary and 
chandra~ Observation log of the BAL QSOs 
\label{chandra_log}}
\tablewidth{0pt}
\tablehead{
\colhead{Object Name} &
\colhead{RA-2000} &
\colhead{Dec-2000} &
\colhead{z} &
\colhead{$D_{\rm L}$\tablenotemark{1}} &
\colhead{$N_{\rm H, gal}$\tablenotemark{2}} &
\colhead{E$_{\rm B-V}$} &
\colhead{TargetID} &
\colhead{T-start\tablenotemark{3}} & 
\colhead{T-stop\tablenotemark{3}} 
& \colhead{MJD} 
& \colhead{$T_{\rm exp}$\tablenotemark{4}}
} 
\startdata
SDSS J073733+392037 & 07 37 33.0 & +39 20 37.4 & 1.7360 &  12499.4 &  5.84  & 0.046 & 13346 & 2011-11-14 07:54  & 2011-11-14 08:54 & 55879.3500 &  1539 \\
IRAS 07598+6508     & 08 04 33.1 & +64 59 49.0 & 0.1483 &    659.6 &  4.17  & 0.047 & 11850 & 2010-06-18 08:11  & 2010-06-18 10:53 & 55365.3958 &  6684 \\
PG 0946+301         & 09 49 41.1 & +29 55 19.0 & 1.2235 &   8081.1 &  1.71  & 0.016 & 11854 & 2010-01-11 08:09  & 2010-01-11 10:30 & 55207.3889 &  6518 \\
PG 1001+054         & 10 04 20.1 & +05 13 00.5 & 0.1611 &    722.4 &  1.83  & 0.014 & 11852 & 2010-01-11 06:23  & 2010-01-11 08:09 & 55207.3021 &  1597 \\
PG 1004+130         & 10 07 26.1 & +12 48 56.1 & 0.2406 &   1132.1 &  3.56  & 0.034 & 05606 & 2005-01-05 16:55  & 2005-01-06 04:37 & 53375.9479 & 41064 \\
PG 1115+080         & 11 18 16.9 & +07 45 58.2 & 1.7355 &  12494.9 &  3.57  & 0.036 & 11857 & 2010-02-01 01:57  & 2010-02-01 06:40 & 55228.1806 & 14582 \\
Mkn 231             & 12 56 14.2 & +56 52 25.2 & 0.0422 &    174.3 &  0.96  & 0.008 & 11851 & 2010-07-11 01:51  & 2010-07-11 03:52 & 55388.1188 &  4782 \\
IRAS 14026+4341     & 14 04 38.8 & +43 27 07.4 & 0.3233 &   1591.4 &  1.15  & 0.010 & 11855 & 2010-07-28 14:41  & 2010-07-28 16:56 & 55405.6583 &  6684 \\
SDSS 1437-0147      & 14 37 48.3 & -01 47 10.7 & 1.311  &   8809.6 &  3.21  & 0.039 & 11851 & 2010-07-11 01:51  & 2010-07-11 03:52 & 55388.1194 &  4782 \\
PG 1700+518         & 17 01 24.8 & +51 49 20.0 & 0.2920 &   1413.8 &  2.26  & 0.030 & 11853 & 2010-06-26 19:51  & 2010-06-26 22:26 & 55373.8819 &  6684 \\
PG 2112+059         & 21 14 52.6 & +06 07 42.5 & 0.466  &   2454.5 &  6.09  & 0.080 & 03011 & 2002-09-01 06:35  & 2002-09-01 23:45 & 52518.6528 & 56868 \\
\enddata
\tablenotetext{1}{The luminosity distance $D_{\rm L}$ was determined using the cosmology calculator by \citet{wright06} 
assuming a Hubble constant $H_{\rm 0}$ = 75 km s$^{-1}$ Mpc$^{-1}$ and is given in units of Mpc}
\tablenotetext{2}{Galactic Column density by \citet{kalberla05} given in units of $10^{20}$ cm$^{-2}$.}
\tablenotetext{3}{Exposure time given in s}
\tablenotetext{4}{Start and end times are given in UT}
\end{deluxetable}

\begin{deluxetable}{lcccccrrrrrrr}
\rotate
\tabletypesize{\tiny}
\tablecaption{\swift~ Observation log of the BAL QSOs
\label{swift_log}}
\tablewidth{0pt}
\tablehead{
\colhead{Object Name} &
\colhead{ObsID} &
\colhead{Segment} &
 \colhead{T-start\tablenotemark{1}} & 
\colhead{T-stop\tablenotemark{1}} 
& \colhead{MJD} 
& \colhead{$\rm T_{\rm XRT}$\tablenotemark{2}} 
& \colhead{$\rm T_{\rm V}$\tablenotemark{2}} 
& \colhead{$\rm T_{\rm B}$\tablenotemark{2}} 
& \colhead{$\rm T_{\rm U}$\tablenotemark{2}} 
& \colhead{$\rm T_{\rm UVW1}$\tablenotemark{2}} 
& \colhead{$\rm T_{\rm UVM2}$\tablenotemark{2}} 
& \colhead{$\rm T_{\rm UVW2}$\tablenotemark{2}} 
} 
\startdata
SDSS J073733+392037 &  39544 & 001 & 2011-08-29 09:41 & 2011-08-22 22:39 & 55795.6319 &  3311 & \nodata & \nodata & \nodata & \nodata &  3294 & \nodata \\
                    &        & 002 & 2012-01-15 03:45 & 2012-01-15 03:50 & 55941.1576 &   235 & \nodata & \nodata & \nodata & \nodata & \nodata & 254 \\
		    &        & 003 & 2012-03-22 05:10 & 2012-03-22 05:20 & 56008.2188 &   584 & \nodata & \nodata & 582 & \nodata & \nodata & \nodata \\ 
IRAS 07598+6508     & 80519  & 001 & 2013-10-29 21:07 & 2013-10-29 23:01 & 56594.9132 &  2093 &  84 &  84 &  81 &   169 &  1296 &  338 \\
PG 0946+301         & 08523  & 001 & 2013-11-09 23:05 & 2013-11-09 23:32 & 56605.9708 &  1606 & 131 & 131 & 131 &   261 &   397 &  522 \\
                    &        & 002 & 2013-11-10 00:41 & 2013-11-10 00:58 & 56606.0347 &  1014 &  85 &  85 &  85 &   171 &   227 &  342 \\
PG 1001+054         & 80521  & 001 & 2013-06-28 08:26 & 2013-06-28 10:22 & 56471.3889 &  1591 &  \nodata & 95 & 95 & 1256 & \nodata & 247 \\
PG 1004+130         & 80031  & 001 & 2012-10-29 08:12 & 2012-10-29 10:04 & 56229.3805 &  1975 & \nodata & \nodata & \nodata & \nodata & \nodata & 1966 \\
                    & 33077  & 001 & 2014-01-10 00:36 & 2014-01-10 08:34 & 56667.1910 &  1838 &  156 &  156 &  156 &  312 &  442 &  626 \\
PG 1115+080         & 90073  & 008 & 2010-02-19 10:19 & 2010-02-20 15:42 & 55247.0729 & 18832 & \nodata & \nodata & \nodata & 1650 & \nodata & 1305 \\
                    &        & 010 & 2010-03-26 02:32 & 2010-03-26 23:36 & 55281.5431 & 19102 & \nodata & \nodata & \nodata & 1569 & \nodata & 1335 \\  
                    & 33106  & 001 & 2014-01-17 16:30 & 2014-01-14 18:22 & 56674.7257 &  2168 &  179 &  179 &  179 &  358 & 487 & 718 \\
Mkn 231             & 32530  & 001 & 2012-08-26 04:45 & 2012-08-26 11:29 & 56165.3368 &  2303 & \nodata & \nodata & \nodata & \nodata & \nodata & 2291 \\
                    & 80091  & 001 & 2013-05-09 13:02 & 2013-05-09 13:21 & 56421.5500 &  1124 & \nodata & \nodata & \nodata & \nodata & \nodata & 1121 \\
                    & 32530  & 002 & 2014-01-10 19:53 & 2014-01-10 21:43 & 56667.8667 &  2437 &  202 &  202 &  202 &  404 &  556 &  810 \\ 
IRAS 14026+4341     & 80520  & 001 & 2013-11-10 21:54 & 2013-11-10 23:43 & 56606.9493 &  1868 &  87 &  87 &  87 &  175 & 1072 & 351 \\
SDSS 1437-0147      & 91050  & 001 & 2012-03-08 09:14 & 2012-03-08 12:28 & 55994.4521 &   250 & \nodata & \nodata & \nodata & \nodata & 251 & \nodata \\
                    &        & 002 & 2012-03-08 09:16 & 2012-03-08 12:37 & 55994.4556 &  1481 & \nodata & \nodata & \nodata & 284 & \nodata & \nodata \\
                    & 33097  & 002 & 2014-01-18 12:22 & 2014-01-18 14:06 & 56675.5521 &  1029 &  35 & 158 & 158 &  315 &  103 & 215  \\ 
                    &        & 003 & 2014-01-22 12:14 & 2014-01-22 12:31 & 56679.5153 &  1006 &  83 &  83 &  84 &  168 &  231 & 336 \\
PG 1700+518         & 37605  & 001 & 2008-07-06 12:19 & 2008-07-07 20:35 & 54654.1875 &  7364 & 626 & 626 & 626 & 1253 & 1603 & 2506 \\
                    &        & 002 & 2008-07-08 02:28 & 2008-07-08 09:14 & 54655.2438 &  2892 & 347 & 347 & 347 &  694 &  622 & 1389 \\
		    & 80179  & 001 & 2012-09-22 21:18 & 2012-09-22 21:36 & 56192.8916 &  1064 & \nodata & \nodata & 1063 & \nodata & \nodata & \nodata \\ 
PG 2112+059         & 33320  & 001 & 2014-06-19 14:31 & 2014-06-19 19:25 & 56827.7069 &  1284 & \nodata & \nodata &  120 & 1140 & \nodata & \nodata \\
                    &        & 002 & 2014-06-21 00:24 & 2014-06-21 00:30 & 56829.0188 &   331 & \nodata & \nodata & \nodata & \nodata & \nodata & 326 \\
		    &        & 003 & 2014-06-25 04:55 & 2014-06-25 21:20 & 56833.5472 &  1638 & \nodata & \nodata & \nodata & \nodata & \nodata & 1634 \\         
                    &        & 004 & 2014-07-01 14:44 & 2014-07-01 16:28 & 55379.6493 &   747 & \nodata & \nodata & \nodata & 735 & \nodata & \nodata \\
                    &        & 005 & 2014-07-03 07:48 & 2014-07-03 07:58 & 56841.3285 &   587 &  48     &   48    & 48      & 97  & 130 & 194 \\
                    &        & 006 & 2014-07-08 22:34 & 2014-07-08 22:37 & 56846.9410 &   152 & \nodata &   19    & 43      & 85 & \nodata & \nodata \\
\enddata

\tablenotetext{1}{Start and end times are given in UT}
\tablenotetext{2}{Observing time given in s}
\end{deluxetable}

\begin{deluxetable}{lcccrrrr}
\rotate
\tabletypesize{\tiny}
\tablecaption{2MASS and WISE fluxes\tablenotemark{1}
\label{ir_res}}
\tablewidth{0pt}
\tablehead{
& \multicolumn{3}{c}{2MASS} & \multicolumn{4}{c}{WISE} \\
\colhead{Object Name} 
& \colhead{J} 
& \colhead{H} 
& \colhead{K} 
& \colhead{W1} 
& \colhead{W2} 
& \colhead{W3} 
& \colhead{W4} 
} 
\startdata
SDSS J073733+392037  & 0.319\plm0.012 & 0.240\plm0.013 & 0.173\plm0.011 & 0.176\plm0.004 & 0.241\plm0.005 & 0.433\plm0.010 & 0.476\plm0.024 \\
IRAS 07598+6508      & 2.496\plm0.075 & 3.731\plm0.094 & 6.182\plm0.109 & 7.378\plm0.158 & 8.086\plm0.143 & 5.873\plm0.076 & 6.071\plm0.090 \\
PG 0946+301          & 0.341\plm0.013 & 0.257\plm0.012 & 0.219\plm0.009 & 0.265\plm0.006 & 0.410\plm0.008 & 0.394\plm0.010 & 0.302\plm0.016 \\
PG 1001+054          & 0.362\plm0.017 & 0.393\plm0.014 & 0.527\plm0.016 & 0.807\plm0.017 & 0.837\plm0.016 & 0.633\plm0.012 & 0.540\plm0.021 \\
PG 1004+130          & 0.806\plm0.023 & 0.635\plm0.025 & 0.673\plm0.020 & 0.715\plm0.015 & 0.765\plm0.014 & 0.955\plm0.016 & 1.121\plm0.038 \\
PG 1115+080          & 0.259\plm0.017 & 0.231\plm0.017 & 0.185\plm0.013 & 0.213\plm0.005 & 0.305\plm0.006 & 0.464\plm0.010 & 0.478\plm0.019 \\
Mkn 231        & 14.880\plm0.280 & 20.566\plm0.346 & 27.600\plm0.433 & 30.110\plm0.872 & 33.400\plm0.747 & 44.961\plm0.416 & 87.351\plm0.895 \\
IRAS 14026+4341      & 0.799\plm0.018 & 0.865\plm0.023 & 1.169\plm0.021 & 1.896\plm0.041 & 2.149\plm0.040 & 2.295\plm0.034 & 2.778\plm0.057 \\
SDSS 1437-0147       & 0.571\plm0.017 & 0.519\plm0.019 & 0.356\plm0.013 & 0.393\plm0.008 & 0.635\plm0.012 & 0.697\plm0.012 & 0.657\plm0.021 \\
PG 1700+518          & 1.140\plm0.024 & 1.159\plm0.035 & 1.490\plm0.027 & 2.124\plm0.044 & 2.341\plm0.041 & 2.249\plm0.027 & 2.592\plm0.046 \\
PG 2112+059          & 0.571\plm0.019 & 0.543\plm0.016 & 0.666\plm0.019 & 1.034\plm0.022 & 1.268\plm0.024 & 0.976\plm0.015 & 1.063\plm0.031 \\
\enddata
\tablenotetext{1}{The fluxes are given in units of $10^{-14}$ W m$^{-2}$, or $10^{-11}$ \erg.}
\end{deluxetable}

\begin{deluxetable}{lclcrcrccc}
\tabletypesize{\tiny}
\tablecaption{Analysis of the Chandra data
\label{chandra_res}}
\tablewidth{0pt}
\tablehead{
\colhead{Object Name} 
& \colhead{MJD} 
& \colhead{Model\tablenotemark{1}} 
& \colhead{\ax} 
& \colhead{$N_{\rm H,z}$\tablenotemark{2}} 
& \colhead{$f_{\rm pc}$} 
& \colhead{$F_{\rm 0.3-10 keV}$\tablenotemark{3}} 
& \colhead{\aox} 
& \colhead{$\alpha_{\rm ox, expected}$\tablenotemark{4}} 
& \colhead{$\Delta$\aox}
} 
\startdata
SDSS J073733+392037  & 55879.3500 & powl    & 0.80$^{+0.42}_{-0.41}$ & --- & --- &  4.33\plm0.80 & 1.80\plm0.07 & 1.66 & +0.14 \\
IRAS 07598+6508      & 55365.3958 & powl    & 1.37$^{+0.91}_{-0.68}$ & --- & --- &  4.45\plm1.21 & 2.34\plm0.10 & 1.47 & +0.87 \\
PG 0946+301          & 55207.3889 & zwa     & 1.09$^{+0.98}_{-0.86}$ & 13.06$^{+9.15}_{-6.96}$ & --- & 3.15\plm0.25 & 1.63\plm0.35 & 1.59 & +0.04 \\
PG 1001+054          & 55207.3021 & powl    & 0.86$^{+1.03}_{-0.99}$ & --- & --- & 0.24\plm0.13 & 2.20\plm0.17 & 1.40 & +0.80 \\
PG 1004+130          & 53375.9479 & zpcfabs & 0.50\plm0.21           & 1.15$^{+0.66}_{-0.72}$ & 0.64$^{+0.10}_{-0.14}$ & 6.63\plm0.15 & 1.86\plm0.06 & 1.49 & +0.37 \\
PG 1115+080          & 55228.1806 & zpcfabs & 1.35$^{+0.40}_{-0.38}$ & 10.65$^{+4.57}_{-4.16}$ & 0.92$^{+0.05}_{-0.06}$ & 7.27\plm0.50 & 1.42\plm0.10 & 1.61 & --0.19 \\
Mkn 231              & 55388.1188 & zpcfabs & 1.00 (fix)             & 4.75$^{+3.23}_{-1.68}$ & 0.90$^{+0.03}_{-0.04}$ & 17.23\plm1.84 & 1.49\plm0.06 & 1.26 & +0.23 \\
IRAS 14026+4341      & 55405.6583 & powl    & 1.00 (fix)             & ---  & --- & 0.06$^{+0.03}_{-0.02}$\tablenotemark{5} & 2.23\plm0.37 & 1.50 & +0.73 \\
SDSS 1437-0147       & 55388.1194 & powl    & 0.68$^{+0.32}_{-0.30}$ & ---  & --- & 12.11\plm0.17 & 1.72\plm0.07 & 1.65 & +0.07 \\
PG 1700+518          & 55373.8819 & powl    & 0.89$^{+1.10}_{-0.92}$ & ---  & --- & 0.15\plm0.01  & 2.42\plm0.12 & 1.52 & +0.90 \\
PG 2112+059          & 52518.6528 & zpcfabs & 0.75\plm0.32           & 5.83$^{+1.75}_{-1.62}$ & 0.82$^{+0.08}_{-0.13}$ & 2.04\plm0.10 & 1.78\plm0.05 & 1.57 & +0.21 \\
\enddata
\tablenotetext{1}{Models: powl = power law model; zwa = absorption at the redshift of the source with power law continuum; 
zpcfabs = partial covering absorber model with power law continuum}
\tablenotetext{2}{Absorption Column density at the redshift of the source given in units of $10^{22}$ cm$^{-2}$.}
\tablenotetext{3}{The absorption corrected 0.3-10 keV flux in the observed frame is given in units of $10^{-16}$ W m$^{-1}$, $10^{-13}$ \erg.}
\tablenotetext{4}{Calculated using the relation given in \citet{grupe10} and the rest-frame k-corrected luminosity density at 2500\AA\ as listed 
in Table\,\ref{swift_res}.}
\tablenotetext{5}{The flux was determined from the ACIS-S count rate of (7.2$^{+3.7}_{-2.8}$)$\times 10^{-4}$ counts s$^{-1}$ assuming a power law model with 
\ax=1.0 and Galactic absorption. }
\end{deluxetable}

\begin{deluxetable}{lcccrrrrrrrcc}
\rotate
\tabletypesize{\tiny}
\tablecaption{Fluxes of the BAL QSOs from the \swift\ observations
\label{swift_res}}
\tablewidth{0pt}
\tablehead{
\colhead{Object Name} &
\colhead{ObsID} &
\colhead{Segment} 
& \colhead{MJD} 
& \colhead{$\rm F_{\rm 0.3-10 keV}$\tablenotemark{1}} 
& \colhead{$\rm F_{\rm V}$\tablenotemark{2}} 
& \colhead{$\rm F_{\rm B}$\tablenotemark{2}} 
& \colhead{$\rm F_{\rm U}$\tablenotemark{2}} 
& \colhead{$\rm F_{\rm UVW1}$\tablenotemark{2}} 
& \colhead{$\rm F_{\rm UVM2}$\tablenotemark{2}} 
& \colhead{$\rm F_{\rm UVW2}$\tablenotemark{2}} 
& \colhead{\auv} 
& \colhead{$\rm l_{\rm 2500\AA}$\tablenotemark{3}} 
} 
\startdata
SDSS J073733+392037 & 39544  & 001 &  55795.6319 &  3.11\plm0.72       & \nodata & \nodata & \nodata & \nodata &  0.45\plm0.02 & \nodata \\
                    &        & 002 &  55941.1576 &                     & \nodata & \nodata & \nodata & \nodata & \nodata & 0.26\plm0.02 & --- & 
		    7.24 \\
		    &        & 003 &  56008.2188 &  \rb{5.63\plm1.35\tablenotemark{4}  }
		           & \nodata & \nodata & 0.94\plm0.04 & \nodata & \nodata & \nodata \\ 
IRAS 07598+6508     & 80519  & 001 &  56594.9132 &  $<$1.03\tablenotemark{5} &  3.35\plm0.13 & 3.84\plm0.12 &  4.43\plm0.22 & 3.49\plm0.20 
                                                                             &  3.31\plm0.10 & 2.51\plm0.12 & 0.39\plm0.01 & 0.154\\
PG 0946+301         & 08523  & 001 &  56605.9708 &                      & 0.45\plm0.04 & 0.53\plm0.02 & 0.53\plm0.03 &  0.42\plm0.03 & 0.33\plm0.02 & 0.22\plm0.01 \\
                    &        & 002 &  56606.0347 &  \rb{1.9$^{+1.9}_{-1.0}$\tablenotemark{4}} & 0.49\plm0.04 & 0.58\plm0.03 & 0.50\plm0.03 & 0.40\plm0.03 
		      & 0.35\plm0.02  &  0.22\plm0.02 & 0.61\plm0.22 & 1.95 \\
PG 1001+054         & 80521  & 001 &  56471.3889 &  $<$2.24\tablenotemark{5} &  \nodata & 0.66\plm0.04 & 0.82\plm0.05 & 0.82\plm0.04 
                                          & \nodata & 0.91\plm0.05 & 0.65\plm0.12 & 0.040 \\
PG 1004+130         & 80031  & 001 &  56229.3805 &  1.91$^{+0.95}_{-0.71}$\tablenotemark{6}   & \nodata & \nodata & \nodata & \nodata & \nodata & 1.47\plm0.06 \\
                    & 33077  & 001 &  56667.1910 &  2.41$^{+1.09}_{-0.48}$ & 1.43\plm0.06 & 1.92\plm0.06 & 2.25\plm0.09 & 1.90\plm0.11 
		                                        & 2.08\plm0.07 & 1.58\plm0.07 & 0.68\plm0.20 & 0.245\\
PG 1115+080         & 90073  & 008 &  55247.0729 & 3.70\plm0.61 & \nodata & \nodata & \nodata & 0.91\plm0.05 & \nodata & 0.64\plm0.03 \\
                    &        & 010 &  55281.5431 & 4.17\plm0.58 & \nodata & \nodata & \nodata & 0.90\plm0.05    & \nodata & 0.64\plm0.03   \\  
                    & 33106  & 001 &  56674.7257 & 5.83\plm1.01 &  0.62\plm0.03 &  0.81\plm0.03 & 1.04\plm0.05 & 0.66\plm0.04 &  & 0.51\plm0.03
		                                   & $-$0.15\plm0.06 & 2.63 \\
Mkn 231             & 32530  & 001 &  56165.3368 &  8.60\plm2.13 & \nodata & \nodata & \nodata & \nodata & \nodata & 0.57\plm0.03 \\
                    & 80091  & 001 &  56421.5500 &  6.85$^{+2.70}_{-2.23}$ & \nodata & \nodata & \nodata & \nodata & \nodata & 0.50\plm0.02 \\
                    & 32530  & 002 &  56667.8667 &  1.70\plm0.42 &  7.17\plm0.17 &  5.26\plm0.10 & 2.32\plm0.09 & 0.87\plm0.08 & 0.57\plm0.02 & 0.51\plm0.03 &
		    3.80\plm0.19 & 2.4$\times 10^{-3}$ \\ 
IRAS 14026+4341     & 80520  & 001 &  56606.9493 &  $<$0.98\tablenotemark{5} &  1.34\plm0.07 &  1.47\plm0.06 & 1.20\plm0.06 & 0.52\plm0.04 
                        & 0.29\plm0.01 & 0.19\plm0.01 & 1.25\plm0.36 & 0.309 \\
SDSS 1437-0147      & 91050  & 001 &  55994.4521 &       & \nodata & \nodata & \nodata & \nodata & 1.74\plm0.07 & \nodata \\
                    &        & 002 &  55994.4556 &  \rb{10.48$^{+6.73}_{-2.42}$}\tablenotemark{4} & 
		          \nodata & \nodata & \nodata & 1.84\plm0.10 & \nodata & \nodata \\
                    & 33097  & 002 &  56675.5521 &       & 1.03\plm0.10 & 1.23\plm0.05 & 1.53\plm0.07 & 1.59\plm0.09 & 1.74\plm0.09 & 1.15\plm0.07 & 
		                0.43\plm0.09 & 5.01 \\ 
                    &        & 003 &  56679.5153 &  \rb{16.21$^{+4.45}_{-2.60}$} & 1.18\plm0.07 & 1.40\plm0.06 & 1.66\plm0.08 & 1.67\plm0.10 & 1.58\plm0.07 & 
		                1.16\plm0.06 & 0.53\plm0.14 \\
PG 1700+518         & 37605  & 001 &  54654.1875 &  0.45$^{+0.18}_{-0.14}$\tablenotemark{6}  & 
                        2.22\plm0.06 & 2.48\plm0.06 & 2.54\plm0.06 & 2.11\plm0.07 & 2.21\plm0.08 & 1.40\plm0.06 & 1.08\plm0.12 \\
                    &        & 002 &  54655.2438 & $<$0.75\tablenotemark{5}  & 
		        2.13\plm0.06 & 2.45\plm0.06 & 2.40\plm0.09 &  1.75\plm0.06 & 2.21\plm0.07 & 1.17\plm0.05 & 0.457 \\
		    & 80179  & 001 &  56192.8916 &  $<$2.01\tablenotemark{5} & \nodata & \nodata & 2.73\plm0.11 & \nodata & \nodata & \nodata \\ 
PG 2112+059         & 33320  & 001 &  56827.7069 &  $2.7^{+1.5}_{-1.2}$\tablenotemark{1} & \nodata & \nodata & 1.87\plm0.09 & 1.88\plm0.10 & \nodata & \nodata 
                    & \nodata  & 0.07 \\
		    &        & 002 &  56829.0188 & $<$19.2\tablenotemark{5} & \nodata & \nodata & \nodata & \nodata & \nodata & 2.04\plm0.10 & \nodata \\
		    &        & 003 &  56833.5472 & $<$3.0\tablenotemark{5} & \nodata & \nodata & \nodata & \nodata & \nodata & 2.00\plm0.09 & \nodata \\
		    &        & 004 &  55379.6493 & $<$6.4\tablenotemark{5} & \nodata & \nodata & \nodata & 1.90\plm0.10 & \nodata & \nodata & \nodata 
		    & 0.08 \\  
                    &        & 005 &  56841.3285 & $<$15.7\tablenotemark{5} & 1.49\plm0.10 & 1.81\plm0.08 & 1.88\plm0.11 & 1.70\plm0.12 & 1.92\plm0.11 &
		    1.73\plm0.13 & 0.89\plm0.10 & 0.08 \\
                    &        & 006 & 56846.9410 & \nodata\tablenotemark{7} & \nodata & 1.97\plm0.12 & 2.04\plm0.11 & 1.85\plm0.13 & \nodata & \nodata &
		    \nodata & 0.06 \\
\enddata

\tablenotetext{1}{Observed 0.3-10 keV flux corrected for Galactic absorption in units of $10^{-16}$ W m$^{-2}$ or $10^{-13}$ \erg.}
\tablenotetext{2}{Observed UVOT fluxes corrected for Galactic reddening in units of $10^{-14}$ W m$^{-2}$ or $10^{-11}$ \erg.}
\tablenotetext{3}{k-corrected rest-frame luminosity density at 2500\AA\
 in units of $10^{24}$ W Hz$^{-1}$ or $10^{31}$  erg s$^{-1}$ Hz$^{-1}$ }
\tablenotetext{4}{Spectral analysis from merged dataset}
\tablenotetext{5}{3$\sigma$ upper limits determent using Bayesian statistics as described in \citet{kraft91}. The count rate was converted into flux using WPIMMS and the
spectral parameters obtained from the \chandra\ observation listed in Table\,\ref{chandra_res}.}
\tablenotetext{6}{The count rate was converted into flux using WPIMMS and the
spectral parameters obtained from the \chandra\ observation listed in Table\,\ref{chandra_res}.}
\tablenotetext{7}{Only 152s exposure, not enough to get meaningful upper limits.}
\end{deluxetable}

\begin{deluxetable}{lccccccccccc}
\rotate
\tabletypesize{\tiny}
\tablecaption{Black hole masses and \lledd
\label{mbh_lledd}}
\tablewidth{0pt}
\tablehead{
\colhead{Object Name} &
\colhead{FW(H$\beta$)\tablenotemark{1}} &
\colhead{FW(MgII)\tablenotemark{1}} &
\colhead{log $L_{\rm 5100\AA}$\tablenotemark{2}} &
\colhead{log $L_{\rm 3000\AA}$\tablenotemark{2}} &
\colhead{log $M_{\rm BH}$\tablenotemark{3}} &
\colhead{log $L_{\rm Edd}$\tablenotemark{2}} &
\colhead{log $L_{\rm bol}$\tablenotemark{4}} &
\colhead{log \lledd} & 
\colhead{FW([OIII])\tablenotemark{1}} 
& \colhead{log([OIII]/H$\beta$)} 
& \colhead{log(FeII/H$\beta$)\tablenotemark{5}}
} 
\startdata
SDSS J073733+392037 & --- & 8470 & --- & 2.51 & 9.98 & 4.07 & 4.20 &  +0.12 & --- & --- & ---  \\
IRAS 07598+6508     & 3400 & --- & 1.20 & --- & 8.58 & 2.68 & 2.00 & --0.68 & 1050 & --1.31 & +0.42  \\
PG 0946+301         & --- & 4180 & --- & 2.35 & 9.28 & 3.38 & 2.49 & --0.89 & --- & --- & --- \\
PG 1001+054         & 1500 & --- & 0.54 & --- & 7.43 & 1.53 & 1.53 &  +0.00 & 1200 & --0.94 & --0.03 \\
PG 1004+130         & 8300 & --- & 1.25 & --- & 9.27 & 3.37 & 2.24 & --1.13 & 1100 & --0.63 & --0.42 \\
PG 1115+080         & --- & 4215 & --- & 2.40 & 9.31 & 3.41 & 4.39 &  +0.96 & --- & --- & --- \\
Mkn 231             & 6530 & --- & 0.47 & --- & 8.78 & 2.88 & 0.92 & --1.96 & 980 & --0.93 & +0.32  \\
IRAS 14026+4341     & 2880 & --- & 1.58 & --- & 8.62 & 2.72 & 2.27 & --0.61 & 530 & --2.35 & +0.28 \\
SDSS 1437-0147      & --- & 4770 & --- & 2.42 & 9.43 & 3.53 & 4.04 &  +0.51 & --- & --- & --- \\
PG 1700+518         & 2440 & --- & 1.69 & --- & 8.53 & 2.63 & 2.38 & --0.25 & 720 & --1.23 & --0.01 \\
PG 2112+059         & 3010 & --- & 1.48 & --- & 8.07 & 2.17 & 2.73 &  +0.56 & 1200 & --0.56 & --0.09 \\ 
\enddata
\tablenotetext{1}{All Full Width at Half Maximum are given in units of km s$^{-1}$.}
\tablenotetext{2}{Luminosities are given in units of 10$^{37}$ W, or 10$^{44}$ erg s$^{-1}$ }
\tablenotetext{3}{Black hole masses in units of \msun, determined using the relations given in \citet{vestergaard06} and \citet{vestergaard09}.}
\tablenotetext{4}{Bolometric Luminosities are given in units of 10$^{37}$ W, or 10$^{44}$ erg s$^{-1}$ and are determined between rest-frame
1$\mu$m and 10 keV.}
\tablenotetext{5}{The FeII flux is measured in the blue FeII blend between 4430 to 4700\AA,same as it has been used in \citet{grupe10}.}
\end{deluxetable}

\begin{deluxetable}{lcccccccccccc}
\rotate
\tabletypesize{\tiny}
\tablecaption{Mean, standard deviation, and median of whole AGN sample including BAL QSOs,  NLS1s\tablenotemark{1} and 
BAL QSOs
\label{agn_statistics}}
\tablewidth{0pt}
\tablehead{
& \multicolumn{4}{c}{All AGN} & \multicolumn{4}{c}{NLS1s} & \multicolumn{4}{c}{BAL QSOs} \\
\colhead{Property} 
& \colhead{Mean}
& \colhead{SD\tablenotemark{2}}
& \colhead{Median} 
& \colhead{\# of AGN}
& \colhead{Mean}
& \colhead{SD\tablenotemark{2}}
& \colhead{Median} 
& \colhead{\# of NLS1s}
& \colhead{Mean}
& \colhead{SD\tablenotemark{2}}
& \colhead{Median} 
& \colhead{\# of BAL QSOs}
} 
\startdata
\ax                & 1.327 & 0.509 & 1.250 & 123 & 1.661 & 0.470 & 1.590 & 54 & 0.936 & 0.266 & 0.890 & 11 \\
\auv               & 0.869 & 0.772 & 0.690 & 122 & 0.878 & 0.762 & 0.690 & 54 & 0.963 & 1.070 & 0.665 & 10 \\
\aox               & 1.385 & 0.247 & 1.340 & 123 & 1.386 & 0.228 & 1.400 & 54 & 1.899 & 0.345 & 1.800 & 11 \\
$\Delta$\aox       & 0.031 & 0.211 & 0.009 & 119 & 0.059 & 0.201 & 0.009 & 52 & 0.379 & 0.381 & 0.230 & 11 \\
log FWHM(H$\beta$)\tablenotemark{3} & 3.365 & 0.263 & 3.382 & 119 & 3.124 & 0.120 & 3.161 & 54 & 3.539 & 0.253 & 3.491 & 7 \\
log FWHM([OIII])\tablenotemark{3} & 2.608 & 0.259 & 2.607 & 119 & 2.610 & 0.264 & 2.574 & 54 & 2.971 & 0.132 & 3.021 & 7 \\
log [OIII]/H$\beta$ & --0.439 & 0.513 & --0.482 & 119 & --0.507 & 0.498 & --0.569 & 54 & --1.136 & 0.603 & --0.940 & 7 \\
log FeII/H$\beta$   & --0.178 & 0.323 & --0.161 & 119 & --0.001 & 0.264 & --0.023 & 54 & +0.067 & 0.292 & --0.010 & 7 \\
log $M_{\rm BH}$\tablenotemark{4} & 7.458 & 0.846 & 7.430 & 123 & 6.805 & 0.559 & 6.844 & 54 & 8.844 & 0.710 & 8.780 & 11 \\
log \lledd & --0.473 & 0.778 & --0.366 & 123 & --0.014 & 0.613 & 0.000 & 54 & --0.306 & 0.854 & --0.250 & 11 \\
log $\dot{M}$\tablenotemark{5} & --0.649 & 1.057 & --0.700 & 123 & --0.878 & 0.933 & --0.800 & 54 & +0.854 & 1.116 & +0.580 & 11 \\
\enddata
\tablenotetext{1}{NLS1s means that we made a cut simply at 2000 km s$^{-1}$.}
\tablenotetext{2}{Standard deviation SD}
\tablenotetext{3}{FWHM are measured in km s$^{-1}$}
\tablenotetext{4}{Black hole masses are given in units of solar masses, $M_{\rm sun}$.}
\tablenotetext{5}{The mass accretion rate $\dot{M}$ is given in units of solar masses per year.}

\end{deluxetable}

\begin{deluxetable}{lcccccccc}
\tabletypesize{\scriptsize}
\tablecaption{Results from the Principal Component Analysis of the whole sample including the low-redshift BAL QSOs (118 AGN in total).
\label{pca_results}}
\tablewidth{0pt}
\tablehead{
\colhead{Property} 
& \colhead{EV 1}
& \colhead{EV 2} 
& \colhead{EV 3}
& \colhead{EV 4}
& \colhead{EV 5} 
& \colhead{EV 6}
& \colhead{EV 7}
& \colhead{EV 8}
} 
\startdata
Proportion of Variance & 0.3787 & 0.2296 & 0.1240 & 0.1034 & 0.07336 & 0.05541 & 0.03116 & 0.00445 \\
Cumulative Proportion  & 0.3787 & 0.6083 & 0.7322 & 0.8356 & 0.90898 & 0.96439 & 0.99555 & 1.00000 \\
\\
\ax                 & -0.2732 &  0.4948  &  0.2886 &  -0.0355 &  0.2338	& -0.6158 &  0.4022 &  0.0174 \\
\auv                &  0.1994 &  0.2282  & -0.8319 &   0.1878 &  0.0179	& -0.3804 & -0.1182 & -0.1482 \\
\aox                & -0.4958 & -0.2918  & -0.0266 &   0.2369 &  0.2309	&  0.0369 &  0.0167 & -0.7465 \\
$\Delta$ \aox       & -0.4615 & -0.2024  & -0.1780 &   0.4637 &  0.3057	&  0.0362 & -0.0454 &  0.6344 \\
log FWHM(H$\beta$)  &  0.2254 & -0.6027  & -0.1643 &  -0.1103 &  0.0054	& -0.1854 &  0.7128 &  0.0652 \\
log FWHM([OIII)     & -0.3070 & -0.1361  & -0.2423 &  -0.8145 &  0.3149	& -0.0666 & -0.2284 &  0.0963 \\
log [OIII]/H$\beta$ &  0.4000 &  0.2054  & -0.0049 &   0.0407 &  0.7904	&  0.3798 &  0.1511 & -0.0663 \\
log FeII/H$\beta$   & -0.3489 &  0.3912  & -0.3265 &  -0.1227 & -0.2728	&  0.5389 &  0.4886 &  0.0048 \\
\enddata

\end{deluxetable}

\clearpage

\begin{appendix}

\section{Additional soft X-ray selected AGN}

In addition to the \swift\ observations of the 92 AGN already published in \citet{grupe10} we added the
remaining AGN in the bright ROSAT AGN sample in \citet{grupe01, grupe04} to our sample. These observations were
performed by \swift\ after publishing the initial sample of 92 objects \citep{grupe10}. These AGN are listed
in Table\,\ref{agn_list_add}. This table contains the coordinated, redshift, H$\beta$ and [OIII] line widths,
black hole masses, Eddington ratios, [OIII]/H$\beta$ and FeII/H$\beta$ flux ratios, and \ax, \auv\ and \aox\
spectral slopes. A description of the \swift\ observations of these additional AGN will be published in a
separate paper. 
Note that this table also contains the information on WPVS 007.

\begin{deluxetable}{lcccrrccccccc}
\rotate
\tabletypesize{\tiny}
\tablecaption{List of additional bright soft X-ray selected AGN observed by \swift
\label{agn_list_add}}
\tablewidth{0pt}
\tablehead{
\colhead{Object Name} &
\colhead{RA-2000} &
\colhead{Dec-2000} &
\colhead{z} &
\colhead{FW(H$\beta$)\tablenotemark{1}} &
\colhead{FW([OIII])\tablenotemark{1}} 
& \colhead{log([OIII]/H$\beta$)} 
& \colhead{log(FeII/H$\beta$)\tablenotemark{2}}
& \colhead{log $M_{\rm BH}$\tablenotemark{3}} 
& \colhead{log \lledd} 
& \colhead{\ax} 
& \colhead{\auv} 
& \colhead{\aox} 
} 
\startdata
WPVS 007            & 00 39 15.2 & --51 17 02  & 0.029 & 1620 &  320 & --0.74 &  +0.20 & 6.60 & --0.52 & 
                       2.65\tablenotemark{4} & 0.82\tablenotemark{4} & 1.89\tablenotemark{4}  \\
IRAS 01267--2157    & 01 29 10.7 & --21 41 57  & 0.039 & 2530 &  360 & --0.50 & --0.22 & 7.64 & --0.47 & 1.24 & --0.02 & 1.33 \\
H 0439--271         & 04 41 22.5 & --27 08 20  & 0.084 & 2550 &  340 &  +0.29 & --0.60 & 7.58 & --0.43 & 1.00 &  +2.31 & 1.04 \\
1 ES 0614--584      & 06 15 49.6 & --58 26 06  & 0.057 & 1080 &  200 & --0.89 &  +0.00 & 6.23 &  +0.79 & 2.19 &  +1.45 & 1.24 \\
RX J1019.0+3752     & 10 19 00.5 &  +37 52 41  & 0.135 & 6200 &  800 & --0.19 & --0.46 & 8.23 & --1.30 & 0.74 &  +1.10 & 1.09 \\
RX J1034.6+3938     & 10 34 38.6 &  +39 38 28  & 0.044 &  700 &  340 &  +0.22 &  +0.14 & 5.82 &  +0.87 & 2.74 &  +1.83 & 1.29 \\
PG 1115+407         & 11 18 30.4 &  +40 25 55  & 0.154 & 1740 &  340 & --1.00 & --0.14 & 7.52 &  +0.07 & 1.45 &  +0.45 & 1.43 \\
Z 1136+3412         & 11 39 13.9 &  +33 55 51  & 0.033 & 1450 &  190 & --0.52 &  +0.01 & 6.27 & --0.30 & 1.39 &  +0.17 & 1.24 \\
Was 26              & 11 41 16.2 &  +21 56 21  & 0.063 & 2200 &  220 &  +0.58 & --0.22 & 7.26 & --0.33 & 1.08 &  +0.70 & 1.32 \\
Mkn 1310            & 12 01 14.4 & --03 40 41  & 0.019 & 3000 &  150 &  +0.11 & --0.21 & 6.67 & --1.52 & 0.85 &  +1.91 & 1.15 \\
Mkn 771             & 12 32 03.6 &  +20 09 30  & 0.064 & 3200 &  210 & --0.55 & --0.38 & 7.40 & --0.57 & 1.12 &  +0.79 & 1.38 \\
CBS 150             & 12 33 41.7 &  +31 01 03  & 0.290 & 1350 &  460 & --0.60 & --0.09 & 7.36 &  +0.95 & 1.58 &  +0.42 & 1.50 \\ 
IRAS 12397+3333     & 12 42 10.6 &  +33 17 03  & 0.044 & 1640 &  510 &  +0.27 &  +0.25 & 6.66 &  +0.28 & 1.47 &  +3.50 & 0.83 \\
PG 1244+026         & 12 46 35.2 &  +02 22 09  & 0.049 &  830 &  330 & --0.06 &  +0.11 & 6.07 &  +0.82 & 1.75 &  +0.78 & 1.31 \\
PG 1322+659         & 13 23 49.5 &  +65 41 48  & 0.168 & 3100 &  130 & --1.00 & --0.37 & 8.16 & --0.40 & 1.70 &  +0.43 & 1.54 \\
QSO 1421--0013      & 14 24 03.6 & --00 26 58  & 0.151 & 1500 &  630 & --0.85 & --0.16 & 7.26 &  +0.63 & 2.16 &  +0.47 & 1.52 \\
SBS 1527+564        & 15 29 07.5 &  +56 16 07  & 0.100 & 2760 &  200 &  +0.52 & --0.16 & 7.15 & --0.38 & 0.82 &  +0.65 & 1.33 \\
ESO 404-G029        & 22 07 45.0 & --32 35 01  & 0.063 & 6100 &  280 & --0.26 & --0.38 & 8.10 & --2.00 & 0.77 &  +2.06 & 1.18 \\
NGC 7214            & 22 09 07.6 & --27 48 36  & 0.023 & 4700 &  530 & --0.24 & --0.06 & 7.54 & --1.70 & 0.93 &  +1.57 & 1.26 \\
PKS 2227--399       & 22 30 40.3 & --39 42 52  & 0.318 & 3710 &  380 &  +0.56 & --0.41 & 7.37 & --0.05 & 0.69 &  +0.03 & 1.24 \\  
\enddata
\tablenotetext{1}{All Full Width at Half Maximum are given in units of km s$^{-1}$.}
\tablenotetext{2}{The FeII flux is measured in the blue FeII blend between 4430 to 4700\AA,same as it has been used in \citet{grupe10}.}
\tablenotetext{3}{Black hole masses in units of \msun, determined using the relations given in \citet{vestergaard06} and \citet{vestergaard09}.}
\tablenotetext{4}{Based on the \swift\ observation of WPVS 007 on 2009-September-17 as reported in \citet{grupe13}.}
\end{deluxetable}

\end{appendix}

\end{document}